\documentclass[aps,prb,twocolumn,nofootinbib,citeautoscript,10pt]{revtex4-2}  
\usepackage{amsmath}
\usepackage{tabularx,graphicx}
\usepackage{epstopdf}
\usepackage{graphicx}
\usepackage{latexsym}
\usepackage{amssymb}
\usepackage{color, colortbl}
\usepackage{psfrag}
\usepackage{bbm}
\usepackage{bm}
\usepackage{titlesec}
\usepackage{dsfont}
\usepackage{feynmp}
\usepackage{slashed}
\usepackage{multirow}

\usepackage[tight]{subfigure}

\usepackage[papersize={8.5in,11in}]{geometry}

\usepackage{color}
\definecolor{darkblue}{rgb}{0.,0.,0.4}
\definecolor{darkred}{rgb}{0.5,0.,0.}
\definecolor{BlueViolet}{RGB}{138,43,226}
\definecolor{SkyBlue}{RGB}{30,144,255}
\definecolor{DarkGreen}{RGB}{0,100,0}
\usepackage[pdftex,colorlinks=true,linkcolor=darkblue,citecolor=blue,urlcolor=darkred]{hyperref}

\newcommand{\beq}{\begin{equation}}
\newcommand{\eeq}{\end{equation}}

\renewcommand{\vec}[1]{{\mathbf{#1}}}

\def\be{\begin{eqnarray}}
\def\ee{\end{eqnarray}}
\def \be{\begin{equation}}
\def \ee{\end{equation}}
\def \bea{\begin{eqnarray}}
\def \eea{\end{eqnarray}}
\def \nn{\nonumber \\}

\begin{document}

\title{Critical Fermi surfaces in generic dimensions arising from transverse gauge field interactions}

\author{Ipsita Mandal}

\affiliation{Faculty of Science and Technology, University of Stavanger, 4036 Stavanger, Norway}
\affiliation{Nordita, Roslagstullsbacken 23, SE-106 91 Stockholm, Sweden}

\begin{abstract}
We study critical Fermi surfaces in generic dimensions arising from coupling finite-density fermions with transverse gauge fields, by applying the dimensional regularization scheme developed previously [\emph{Phys. Rev. B 92, 035141 (2015)}]. We consider the cases of $U(1)$ and $U(1)\times U(1)$ transverse gauge couplings, and extract the nature of the renormalization group (RG) flow fixed points as well as the critical scalings.
Our analysis allows us to treat a critical Fermi surface of a generic dimension $m$ perturbatively in an expansion parameter $\epsilon =\left (2-m \right ) /\left (m+1 \right).$
One of our key results is that although the two-loop corrections do not alter the existence of an RG flow fixed line for certain $U(1)\times U(1)$ theories, which was identified earlier for $m=1$ at one-loop order, the third-order diagrams do. However, this fixed line feature is also obtained for $m>1$, where the answer is one-loop exact due to UV/IR mixing.
\end{abstract}
\maketitle

\tableofcontents

\maketitle

\section{Introduction}

Metallic states that lie beyond the framework of Landau Fermi liquid theory are often dubbed non-Fermi liquids. 
A finite density of of nonrelativistic fermions coupled to a transverse $U(1)$ gauge field has been known as an example of a non-Fermi liquid. This model was first studied by Holstein, Norton, and Pincus \cite{holstein}, whose original motivation was to understand the effects of the electromagnetic field coupled to a metal. However, it was realized that the same field theory applies to the case of fermions coupled to abelian / non-abelian emergent gauge field(s) in various scenarios like certain quantum spin liquids \cite{PhysRevResearch.2.013072}, normal state of cuprate superconductors \cite{baskaran,larkin,PhysRevLett.63.680,leenag,blok,ubbens}, and the compressible quantum Hall systems at $1/2$-filling \cite{nayak1,Chakravarty}.
Such a fictitious transverse gauge field has the gauge coupling value of the order unity (and not the value $1/137$  of the fine structure constant of the electromagnetic forces) and consequently, gives rise to a strongly correlated system.
It is a theoretically challenging task to study such systems, and consequently there have been intensive efforts dedicated to building a framework to understand them  \cite{holstein,reizer,leenag,HALPERIN,polchinski,ALTSHULER,Chakravarty,eaKim,nayak,nayak1,lawler1,SSLee,
metlsach1,metlsach,chubukov1,Chubukov,mross,Jiang,ips2,ips3,Shouvik1,Lee-Dalid,shouvik2,ips-uv-ir1,ips-uv-ir2,ips-subir,ips-sc,ips-c2,andres1,andres2,Lee_2018,ips-fflo}. They are also referred to as critical Fermi surface states, as the breakdown of the Fermi liquid theory is brought about by the interplay between the soft fluctuations of the Fermi surface and some gapless bosonic fluctuations. These bosonic degrees of freedom can be massless scalar bosons, or the transverse components of gauge fields.
A similar situation also arises in semimetals, where instead of a Fermi surface, there is a Fermi node interacting with long-ranged (unscreened) Coulomb potential which gives rise to a non-Fermi liquid behaviour \cite{abrikosov,moon-xu,rahul-sid,ips-rahul}.
Since the quasiparticles are destroyed, there is no obvious perturbative parameter in which one can carry out a controlled expansion, which would ultimately enable us to extract the universal properties.

 In this paper, we consider the case when Fermi surfaces are coupled with emergent gauge fields \cite{Chakravarty,MOTRUNICH,LEE_U1,PALEE,MotrunichFisher,nayak1,mross,ips2,ips3}.
This belongs to the category when the critical boson carries zero momentum, and the
quasiparticles lose coherence across the entire Fermi surface. 
An example when the critical boson with zero momentum is a scalar, is the Ising-nematic critical point \cite{metlsach1,ogankivfr,metzner,delanna,kee,lawler1,rech,wolfle,maslov,quintanilla,yamase1,yamase2,halboth,
jakub,zacharias,eaKim,huh,Lee-Dalid,ips-uv-ir1,ips-uv-ir2,ips-subir,ips-sc}. 
There are complementary cases when the critical boson carries a finite momentum.
Examples include the critical points involving spin density wave (SDW), charge density wave (CDW) \cite{metlsach,chubukov1,Chubukov,shouvik2,ips-c2,andres1,andres2},
and the FFLO order parameter \cite{ips-fflo}.

An analytic approach \cite{senshank,Lee-Dalid,ips-uv-ir1,ips-fflo} to deal with non-Fermi liquid quantum critical points is through dimensional regularization, in which the co-dimension of the Fermi surface is increased in order to identify an upper critical dimension $d=d_c$, and subsequently, to calculate the critical exponents in a systematic expansion involving the parameter $\epsilon = d_c - d_{\text{phys}}$
(where $d_{\text{phys}}$ is the actual/physical dimension of the system).
This approach is especially useful, as it allows one to deal with critical Fermi surfaces of a generic
dimension $m$ \cite{ips-uv-ir1,ips-uv-ir2}, representing a system with physical dimensions $d=d_{\text{phys}} = m+1$.
The physical systems have $d_{\text{phys}}$ equal to two or three. Hence, $m$ is equal to one or two for the corresponding systems.

Another approach implements controlled approximation through dynamical tuning, involving an expansion
in the inverse of the number ($N$) of fermion flavours combined with a further expansion
$\varepsilon = z_b-2$, where $z_b$ is the dynamical critical exponent of the boson field  \cite{nayak1,mross}.
This amounts to modifying the kinetic term of a collective mode ($\phi(k)$)
from $k^{2}\,|\phi(k)|^2 $ to $k^{1+\varepsilon}\,|\phi(k)|^2 $. A drawback of this approach is that this modification of the kinetic term leads to nonalayticities in the momentum space, which are equivalent to
nonlocal hopping terms in real space.
Hence in this paper, we will employ the former approach of dimensional regularization, which maintains locality in real space.

The earlier works considering generic values of $d$ and $m$ involved the Ising-nematic order parameter \cite{ips-uv-ir1,ips-uv-ir2}, which represents quantum critical metals near a Pomeranchuk transition, where the critical boson couples to antipodal patches with the same sign of coupling strength \cite{metlsach1}. In contrast, a transverse gauge field couples to the two antipodal patches with opposite signs \cite{SSLee}. 
Here, we will implement the dimensional regularization procedure to determine the low-energy scalings of an $m$-dimensional (with $m\geq 1$) Fermi surface coupled with one or more transverse gauge fields. First we will develop the formalism for a single $U(1)$ gauge field. Then we will extend it to the $U(1) \times U(1)$ case, which can describe a quantum phase transition between a Fermi liquid metal
and an electrical insulator without any Fermi surface (deconfined Mott transition), or that
between two metals that having Fermi surfaces with finite
but different sizes on either side of the transition (deconfined metal-metal transition) \cite{debanjan}.

The paper is organized as follows. In Sec.~\ref{modelu1}, we review the framework for applying dimensional
regularization scheme to access the non-Fermi liquid fixed points perturbatively, and apply it to the case of
a single transverse gauge field. We also compute the renormalization of the $2k_F$
scattering amplitude for the fermions, and the scaling forms of some physical observables.
In Sec.~\ref{modelu2}, we carry out the computations for the scenario of quantum critical transitions involving two different kinds of fermions charged differently under the action of
two transverse gauge fields. We conclude with a summary and an outlook in Sec.~\ref{conclude}.
The details of the one-loop calculations are provided in the Appendix.


\section{Model involving a $U(1)$ transverse gauge field}
\label{modelu1}

\begin{figure}[]
\begin{center}
\subfigure[]{\includegraphics[scale=0.6]{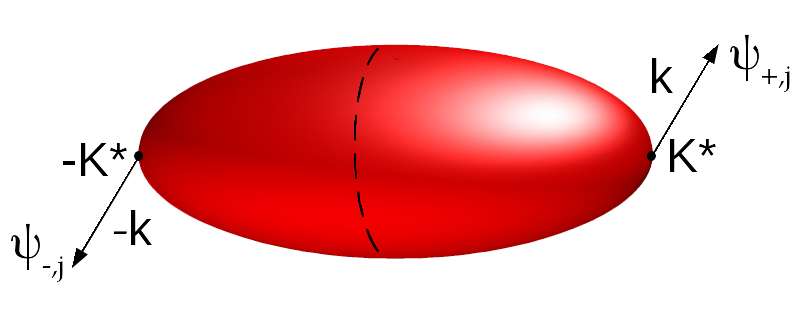} 
\label{fig:FS}} \\
\subfigure[]{\includegraphics[scale=0.6]{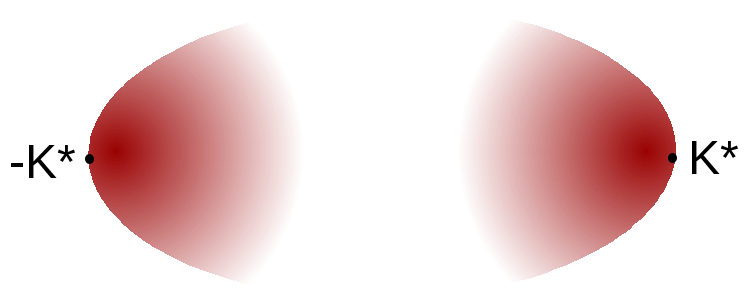} 
\label{fig:patches}} \\
\end{center}
\caption{(a) Schematic of a compact Fermi surface divided into two halves, which are centered at $ K^*$ and $-K^*$ respectively. 
For the two halves, two separate fermionic fields ($\psi_{\pm,j}$) have been introduced.
(b) In the effective action, a compact Fermi surface is approximated by 
two sheets of non-compact Fermi surfaces (approximated by parabolic dispersion to leading order). 
The momentum modes far away from $\pm K^*$ have been suppressed by using a momentum regularization.
}
\end{figure}

\begin{figure}
\centering
\includegraphics[width=0.45 \textwidth]{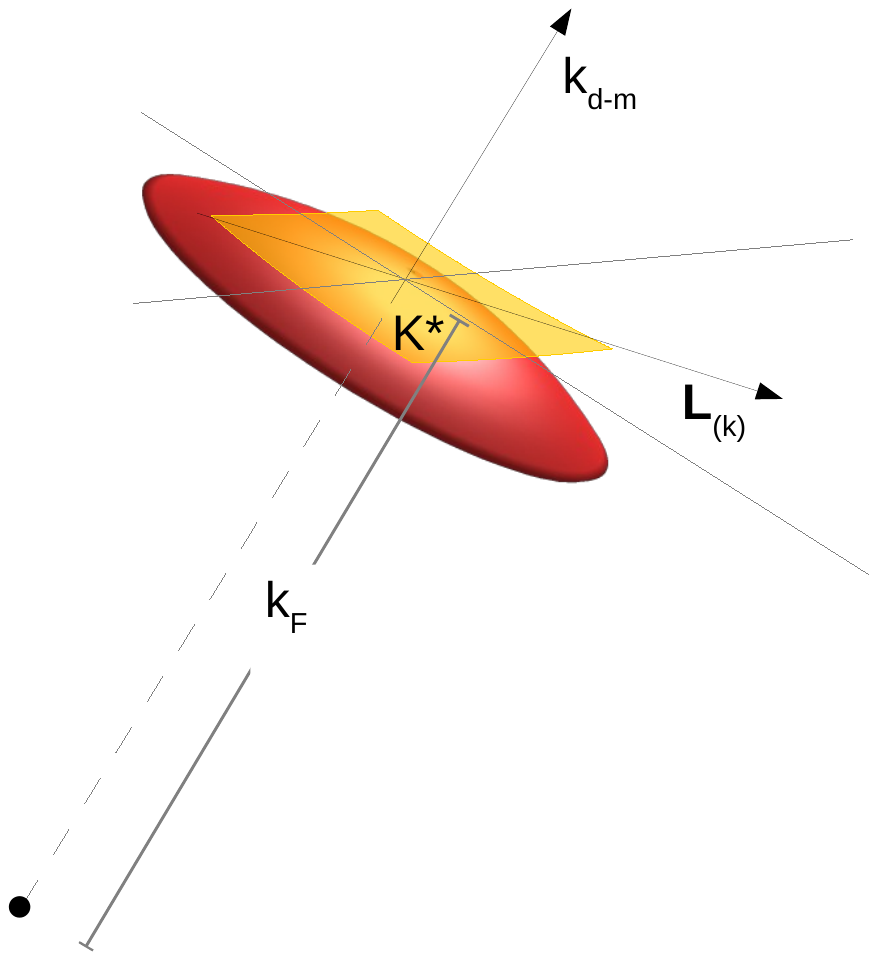} 
\caption{The momentum coordinates on a patch of an $m$-dimensional convex Fermi surface.}
\label{fs-patch}
\end{figure}

We first consider an $m$-dimensional Fermi surface, 
which is coupled to a $U(1)$ transverse gauge field $a $
in $d=(m+1)$ space dimensions. The set-up is identical to Ref.~\cite{ips-uv-ir1}. We review it here for the sake of completeness.
As in earlier works \cite{Lee-Dalid,ips-uv-ir1,ips-uv-ir2}, we want to characterize the resulting non-Fermi liquids through the scaling properties of the fermionic and bosonic Green's functions.
To do so, we focus on one point (say $K^*$) of the Fermi surface
at which the fermion Green's function is defined.
The low energy effective theory involves fermions which are primarily scattered 
along the tangential directions of the Fermi surface, mediated by the critical boson.
We assume the presence of the inversion symmetry, which implies that the
fermions near $K^*$ are most strongly
coupled with fermions near the antipodal point $-K^*$, since
their tangent spaces coincide.
Hence we write down a model including a closed Fermi surface divided into two halves centered at momenta $K^*$  and $-K^*$ respectively. The fermionic fields $\psi_{+,j}$ and $\psi_{-,j}$ represent the corresponding halves, as shown in Fig. \ref{fig:FS}.
In this coordinate system, the minimal Euclidean action that captures the essential description of the low
energy physics is given by \cite{SSLee}:
\begin{align}
S = &   \sum \limits_{p=\pm} \sum_{j=1}^N \int dk\, 
\psi_{p,j}^\dagger (k) 
\left[ \mathrm{i}\,k_0   +  p  \,k_1 +  {\vec L}_{(k)}^2  \right ] \psi_{p,j}(k)
\nn &
 + \frac{1}{2} \int  dk
 \left[ k_0^2 + k_1^2   + {\vec L}_{(k)}^2 \right]
  a^\dagger(k) \, a(k)  \nonumber \\
 &  +  \frac{e}{\sqrt{N}}  \sum_{ p=\pm}  p \sum_{j=1}^N  
\int dk\,dq \, a(q) \,  \psi^\dagger_{p,j}(k+q) 
\, \psi_{ p,j}(k) \, ,
\label{actu1}
\end{align}
where $k =(k_0,k_1, {\vec L}_{(k)})$ is the $(d+1)$-dimensional energy-momentum vector with
$dk \equiv \frac{d^{d+1} k}{(2\pi)^{d+1} }\,,$ and $e$ is the transverse gauge coupling. 
The fermion field
$\psi_{+,j}(k_0,k_i)$ $\left ( \psi_{-,j}(k_0,k_i) \right)$
with flavor $j=1,2,..,N$, frequency $k_0$ and momentum $K_i^*+k_i$ ($-K_i^*+k_i$)
is represented by $\psi_{+,j}(k_0,k_i)$ $\left ( \psi_{-,j}(k_0,k_i) \right)$, with $1 \leq i \leq d$.
The components $k_{1}$ and  ${\vec L}_{(k)} ~\equiv~ (k_{2}, k_{3},\ldots, k_{d})$ 
represent the momentum components perpendicular and parallel to the Fermi surface at $\pm K^*$, respectively. 
We have rescaled the momentum such that the absolute value of the Fermi velocity and the quadratic curvature of the Fermi surface at $\pm K^*$ can be set to one.
An actual physical situation can involve a Fermi surface of an arbitrary shape. Our physical set-up allows us to include such a generic scenario as long as the Fermi surface is locally convex, as the coordinate has been set up with its origin at a particular small patch of the Fermi surface (see Fig.~\ref{fs-patch}).

Due to the fact that the Fermi surface is locally parabolic, the scaling dimensions of $k_1$ and ${\vec L}_{(k)}$ are equal to $1$ and $1/2$ respectively.
For a generic convex Fermi surface, there can be cubic and higher order terms in ${\vec L}_{(k)}$,
but we can ignore them as they irrelevant in the renormalization group (RG) sense. 
Since we have a compact Fermi surface, the range of ${\vec L}_{(k)}$ in $\int dk$
is finite and is set by the size of the Fermi surface. This range is of the order of $\sqrt{k_F}$ in this coordinate system. To ensure this finite integration range,
we will include an exponential cut-off $\exp \left \lbrace- \frac {{\vec{L}}_{(k)}^2}  { \mu \, {\tilde{k}}_F } \right \rbrace$ while using the fermion Green's function in loop integrations, which will capture the compactness of the Fermi surface
in a minimal way without including the details of the shape. This can be made explicit by including the inverse of this factor in the kinetic part of the fermion action.

In order to control the gauge coupling $e$ for a given $m$,
we tune the co-dimension of the Fermi surface \cite{senshank,Lee-Dalid,shouvik2} to determine the upper critical dimension $d=d_c$.
To preserve the analyticity of the theory in momentum space 
(locality in real space) with general co-dimensions, 
we introduce the spinors \cite{Lee-Dalid,shouvik2}
\begin{align}
 \Psi_j^T(k) = \left( 
\psi_{+,j}(k)\quad
\psi_{-,j}^\dagger(-k)
\right) \text{ and } \bar \Psi_j \equiv \Psi_j^\dagger \,\gamma_0\,,
\end{align} 
and write an action that describes the $m$-dimensional Fermi surface
embedded in a $d$-dimensional momentum space:
\begin{align}
\label{actu12}
S  =&   \sum_{j} \int dk\, \bar \Psi_j(k) \,\mathrm{i}
\left[  \vec \Gamma \cdot \vec K  +  \gamma_{d-m} \, \delta_k \right ]
 \Psi_{j}(k) \, \exp \Big \lbrace \frac {{\vec{L}}_{(k)}^2}  { \mu \, {\tilde{k}}_F } \Big \rbrace \nonumber\\
&+
\frac{1}{2} \int  dk \,
  {\vec{L}}_{(k)}^2\,  a^\dagger(k) \, a(k) \nonumber \\
 &+    \frac{  e \, \mu^{x/2} } {\sqrt{N}}  \sum_{j}  
\int dk \,dq  \,
a(q) \, \bar \Psi_{j}(k+q)\,  \gamma_{0}\, \Psi_{j}(k) \,,
\nn x = & \frac{4+m-2d} {2} \,.
\end{align}
Here, $\vec K ~\equiv ~(k_0, k_1,\ldots, k_{d-m-1})$ includes
the frequency and the first $(d-m-1)$ components 
of  the $d$-dimensional momentum vector, ${\vec L}_{(k)} ~\equiv~ (k_{d-m+1}, \ldots, k_{d})$ and $\delta_k =  k_{d-m}+ {\vec{L}}_{(k)}^2$.
In the $d$-dimensional momentum space,
$k_1,..,k_{d-m}$ (${\vec L}_{(k)}$) represent(s) the
$(d-m)$ ($m$) directions perpendicular (tangential) to the Fermi surface.
$\vec \Gamma \equiv (\gamma_0, \gamma_1,\ldots, \gamma_{d-m-1})$ represents the gamma matrices associated with $\vec K$.
Since we are interested in a value of co-dimension $1 \leq d-m \leq 2$, 
we consider only $2 \times 2$ gamma matrices with
$\gamma_0= \sigma_y , \, \gamma_{d-m} = \sigma_x$.
In the quadratic action of the boson, only ${\vec{L}}_{(k)}^2 \, a^\dagger (k)\, a(k)$ 
is kept, because $|\vec K|^2 + k_{d-m}^2$ 
is irrelevant under the scaling 
where $k_0,k_1,..,k_{d-m}$ have dimension $1$
and $k_{d-m+1},..,k_d$ have dimension $1/2$.
In the presence of the $(m+1)$-dimensional rotational symmetry,
all components of $k_{d-m}, ..., k_d$ should be equivalent.
The rotational symmetry of the bare fermion kinetic part in the $(d-m)$-dimensional space spanned by $\vec K$ components is destroyed by the coupling with the gauge boson, as the latter involves the $\gamma_0$
matrix. With this in mind, we will denote the extra (unphysical) co-dimensions by the vector $\tilde{\vec K}$, and the corresponding gamma matrices by $\tilde{\vec \Gamma}$.

Since the scaling dimension of the gauge coupling $e$ is equal to $x/2$, we have
made $e$  dimensionless by using a mass scale $\mu$.
We have also defined a dimensionless parameter for the Fermi momentum, 
$ {\tilde {k}}_F = k_F/\mu$ using this mass scale.
The spinor $\Psi_j$ exhibits an energy dispersion with two bands $E_k =
 \pm \sqrt{ \sum \limits_{i=1}^{d-m-1} k_i^2 + \delta_k^2  } \,,$ and this gives an $m$-dimensional Fermi surface 
embedded in the $d$-dimensional momentum space,
defined by the $d-m$ equations:
$k_i = 0$ for $i=\lbrace 1,\ldots,d-m-1 \rbrace$
and ${ k}_{d-m}  = - {\vec{ L}}_{(k)}^2$.
Basically, the extra ($d-m-1$) directions are gapped out so that the Fermi surface reduces to
a sphere $S^m$ (sphere in an $(m+1)$-dimensional Euclidean space) locally.

When we perform dimensional regularization, the theory implicitly 
has an ultraviolet (UV) cut-off for $\vec K$ and $k_{d-m}$,
which we denote by $\Lambda$.
It is natural to choose $\Lambda = \mu$,
and the theory has two important dimensionless parameters: $e$ and $\tilde k_F = k_F / \Lambda$.
If $k$ is the typical energy 
at which we probe the system, 
the limit of interest is $k \ll \Lambda \ll k_F$. 
This is because  $\Lambda$ sets the largest energy (equivalently, momentum perpendicular to the Fermi surface) 
fermions can have, whereas $k_F$ sets the size of the Fermi surface.
We will consider the RG flow generated 
by changing $\Lambda$ and requiring that low-energy observables are independent of it.
This is equivalent to a coarse-graining procedure of integrating out
high-energy modes away from Fermi surface.
Because the zero energy modes are not integrated out,
$k_F/\Lambda$ keeps on increasing  in the coarse-graining procedure.
We treat $k_F$ as a dimensionful coupling constant 
that flows to infinity in the low-energy limit.
Physically, this describes the fact that 
the size of the Fermi surface, measured in the unit of
the thickness of the thin shell, around the Fermi surface
diverges in the low-energy limit.
This is illustrated in Fig.~\ref{fig:FS}.

\subsection{Dimensional regularization}
\label{dr}

\begin{figure}[h!]
\begin{center}
\subfigure[]{
\includegraphics[scale=0.3]{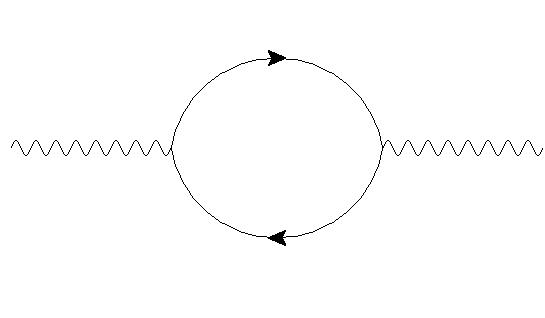} 
\label{fig:bos}
}\\
\subfigure[]{
\includegraphics[scale=0.3]{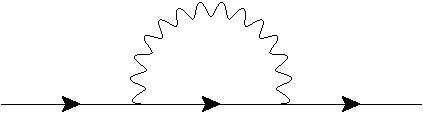} 
\label{fig:ferm}
}\\
\subfigure[]{
\includegraphics[scale=0.3]{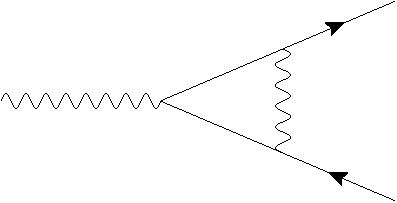} 
\label{fig:vert}
}
\end{center}
\caption{The one-loop diagrams
for (a) the boson self-energy,
(b) the fermion self-energy, and
(c) the vertex correction.
Lines with arrows represent the bare fermion propagator,
whereas wiggly lines in (b) and (c) represent
the dressed boson propagator 
which includes the one-loop self-energy in (a).
}
\label{fig:1loop}
\end{figure}

To gain a controlled approximation of the physics of the critical Fermi surface, 
we fix $m$ and tune $d$ towards a critical dimension $d_c\,,$ 
at which quantum corrections depend logarithmically on $\Lambda$ 
within the range $\Lambda \ll k_F$.
In order to identify the value of $d_c$ as a function of $m$,
we consider the one-loop quantum corrections.

The bare propagator for fermions is given by:
\begin{align}
\label{propf}
G_0 (k) = -\mathrm{i}\, \frac{\vec \Gamma \cdot \vec K +
\gamma_{d-m} \,\delta_k} 
{\vec K^2  + \delta_k^2} \, \times \, \exp \Big \lbrace - \frac {{\vec{L}}_{(k)}^2}  { \mu \, {\tilde{k}}_F } \Big \rbrace \,.
\end{align}
Since the bare boson propagator is independent of $k_{0},..,k_{d-m}$,
the loop integrations involving it are ill-defined,
unless one resums a series of diagrams
that provides a non-trivial dispersion along those directions.
This amounts to rearranging the perturbative expansion 
such that the one-loop boson self-energy is included 
at the `zero'-th order.
The dressed boson propagator includes the one-loop self-energy
(see Fig. \ref{fig:bos}) given by:
\begin{align}
\label{babos}
& \Pi_1 (k) = - e^2 \mu^x
\int  dq\, \text{Tr}
\left[ \gamma_{0}\, G_0 (k+q)\,\gamma_{0}\, G_0 (q) \right ]
\nn & =
- \frac{ \beta(d,m)\,  e^2 \, \mu^x \left(  \mu \, {\tilde{k}}_F  \right )^{\frac{m-1}{2}}  } 
{   |\vec{L}_{(k)}|}  
\nn & \qquad \times \left[ k_0^2 + ( m+1-d)\,{\tilde {\vec K}}^2 \right] 
|\vec K|^{d-m-2} \,,
\end{align}
where
\begin{align}
\label{eqbetad}
\beta(d,m)
 =
\frac{ \pi ^{\frac{4-d}{2}} 
\,\Gamma (d-m) \,\Gamma (m+1-d) }
 {2^{\frac{4d-m-1}{2} }\,\Gamma ^2 \left(\frac{ d-m+2} {2}\right) \Gamma \left(\frac{m+1-d}{2} \right)} \,.
\end{align}
This expression is valid to the leading order in $k/k_F$,
and for $|\vec K|^2/|\vec L_{(k)}|^2, ~\delta_k^2/|\vec L_{(k)}|^2 \ll k_F$
\footnote{The $k_F$-dependence drops out for $m=1$.}.
We provide the details of computation for the expression of $\Pi_1(k) $ in Appendix~\ref{app:oneloopbos}.
For $m>1$, the boson self-energy diverges
in the $k_F \rightarrow \infty$ limit.
This is due to the fact that the Landau damping gets stronger 
for a system with a larger Fermi surface,
as the boson can decay into particle-hole excitations
that encompass the entire Fermi surface for $m>1$.
This is in contrast with the case for $m=1$, 
where a low-energy boson with a given momentum can decay into particle-hole excitations only near the 
isolated patches whose tangent vectors are parallel
to that momentum.
Eq.~(\ref{babos}) is valid when there exists at least one direction that is tangential to the Fermi surface ($m \geq 1$).
Henceforth, we will use the dressed propagator:
\begin{align}
\label{babosprop}
D_1 (k) = \frac{1}{{\vec{L}}_{(k)}^2 - \Pi_1 (k)} \,.  
\end{align}
for any loop calculation.

The next step is to computed the one-loop fermion self-energy $\Sigma_1 (q)$, as shown in Fig.~\ref{fig:ferm}.
Again, the details of the calculation are provided in Appendix~\ref{app:oneloopfer}.
This blows up logarithmically 
in $\Lambda$ at the critical dimension
\begin{align}
d_c(m) = m + \frac{3}{m+1} \,.
\end{align}
The physical dimension is given by $d=d_c(m) - \epsilon$.
In the dimensional regularization scheme, 
the logarithmic divergence in $\Lambda$
turns into a pole in $\frac{1} {\epsilon}$:
\begin{align}
\label{sigmau1}
\Sigma_1(k) = & -\frac{ \mathrm{i} \, 
e^{\frac{2\,(m+1)} {3} } 
\left[ u_0  \,
\gamma_0\,k_0 
+u_1  
\left( {\tilde{\vec \Gamma}} \cdot {\tilde{\vec K}} \right)
 \right]  }
{N \, {\tilde{k}}_F ^{ \frac{(m-1)(2-m) } {6}}
\, \epsilon} 
\nn & + \text{ finite terms }
\end{align}
to the leading order in $k/k_F\,,$ where $u_0\,,u_1\geq 0\,$.
For the cases of interest, we have computed these coefficients numerically to obtain:
\begin{align}
\label{valu}
\begin{cases}
u_0 = 0.0201044 \,, \quad u_1 =1.85988 & \text{ for } m=1 \\
 u_0 = u_1 = 0.0229392 & \text{ for } m=2
\end{cases} \,.
\end{align}

The one-loop vertex correction in Fig.~\ref{fig:vert} is given by (see Appendix~\ref{oneloopvert} for the detailed steps for evaluating the integrals):
\begin{align}
\label{eqvertcor}
\Gamma_{1}(k,0)
& =
-  \frac{e^{\frac{2 \,(m+1)}{3}}\,
 \,u_4 \, \gamma_{0}}
    {N \,\tilde{k}_{F}^{ (m-1) \, (2-m) / 6}\,\epsilon} \,
 \left( \frac{\mu} {| \tilde{\vec K}|} \right)^{\frac{(m+1) \,\epsilon}{3}} 
\left[  {\mathcal F}\bigg ( \frac{|k_0|} { | \tilde{\vec K}|} \bigg) \right]^\epsilon   
\nn &
\quad  + \, \text{finite terms} \,,
\end{align}
where $u_4 \geq 0$ and ${\mathcal F}$ is some dimensionless function of ${|k_0|} /{ | \tilde{\vec K}|}$. Specifically, we have: 
\begin{align}
 u_4 = \begin{cases}
0.0000706373 & \text{ for } m=1 \\
    0 &\text{ for } m=2
\end{cases} \,.   
\end{align}
This is to be contrasted with the Ising-nematic case where it is guaranteed to vanish due to a Ward identity \cite{Lee-Dalid,ips-uv-ir1}.

We can vary the dimension of Fermi surface from $ m = 1$ to $ m=2 $ while keeping $\epsilon$ small, 
thus providing a controlled description for any $m$ between $1$ and $2$.
For a given $m$, we tune $d$ such that $\epsilon =  d_c(m) - d $ is small.
To remove the UV divergences in the $\epsilon \rightarrow 0$ limit,
we add counterterms using the minimal subtraction scheme.
The counterterms take the same form 
as the original local action:
\begin{widetext}
\begin{align}
\label{actcount}
S_{CT}  = &  \sum_{j} \int dk\, \bar \Psi_{j}(k)
\, \mathrm{i} \,\Bigl[ 
A_{0} \,  \gamma_0 \,k_0 + A_{1} \,\tilde{\vec \Gamma} \cdot \tilde{ \vec K} 
+   A_2 \, \gamma_{d-m} \, \delta_k 
 \Bigr] \Psi_{j}(k) \,  \exp \Big \lbrace \frac {{\vec{L}}_{(k)}^2} 
  { \mu \, {\tilde{k}}_F } \Big \rbrace
\nn & + \frac{A_{3}}{2} \int dk\,
 {\vec{L}}_{(k)}^2\,   a^\dagger(k) \, a(k) 
+  A_{4} \frac{ e \, \mu^{x/2} }{\sqrt{N}} 
\sum_{j}  \int dk \, dq  \,
a(q) \,  \bar \Psi_{ j}(k+q) \,\gamma_{0} \, \Psi_{j}(k)  \, ,
\end{align}
\end{widetext}
where 
\begin{align}
A_{\zeta} = 
\sum_{\lambda=1}^\infty \frac{Z^{(\lambda)}_{ \zeta}
(e ,\tilde{k}_F)}{\epsilon^\lambda}  \text{  with }  \zeta=0,1,2,3 , 4\,.
\end{align}
In the mass-independent minimal subtraction scheme,
these coefficients depend only on the scaled coupling $e$, and the scaled Fermi momentum $\tilde{k}_F$. As discussed earlier, we expect $\tilde{k}_F$ to act as another coupling for $m>1$, and hence it must be included in the RG flow equations.
The coefficients can be further expanded in the number of loops
modulo the one-loop self-energy of boson, which 
is already included in Eq.~(\ref{babosprop}).
Note that the $(d-m-1)$-dimensional rotational invariance
in the space perpendicular to the Fermi surface
guarantees that each term in $ \tilde{\vec \Gamma} \cdot \tilde{\vec K}$ 
is renormalized in the same way. Similarly, the sliding symmetry along the Fermi surface
guarantees that the form of $\delta_k$ is preserved. However, $A_0$, $A_1$ and $A_2$ are in general different
due to a lack of the full rotational symmetry in the $(d+1)$-dimensional spacetime.
Note the difference from the Ising-nematic case, where we had $A_0 = A_1$, as the rotational symmetry there involved the full $(d-m)$-dimensional subspace.

Adding the counterterms to the original action,
we obtain the renormalized action
which gives the finite quantum effective action:
\begin{widetext}
\begin{align}
\label{act7}
S_{ren}  = &  \sum_{j} \int d k^B
\, \bar \Psi_{j}^B(k^B) \,\mathrm{i} 
\left[   \gamma_0 \,k_0^B + \tilde{\vec \Gamma} \cdot \tilde{ \vec K}^B
 +  \gamma_{d-m} \delta_{k^B}  \right ] \Psi_{j}^B(k^B) 
\, \exp \left \lbrace \frac {{\vec{L}}_{(k^B)}^2}  {  k_{F^B} } \right \rbrace
+
\frac{1}{2} \int  d k^B \,
 {\vec{L}}_{(k^B)}^2\, { a^B } ^\dagger (k^B) \,\,\,a^B(k^B) \nonumber \\
 &+    \frac{  e^B }{\sqrt{N}}  \sum_{j}  
\int d k^B \, d q^B \,
a^B(q^B) \, \bar \Psi_{j}^B(k^B+q^B) 
\, \gamma_{0} \Psi_{j}^B(k^B) \, ,
\end{align}
\end{widetext}
where
\begin{align}
& k_{0}^B = \frac{Z_0} {Z_2}\,k_0\,,\quad
\tilde{\vec K}^B =   \frac{Z_1} {Z_2} \, \tilde{\vec K} \, , \quad
k_{d-m}^B =  k_{d-m} \, , 
\nn & {\vec{L}}_{(k^B)}  =  {\vec{L}}_{(k)} \,, \quad
\Psi_j^B(k^B)  =   Z_{\Psi}^{\frac{1}{2}}\, \Psi_j(k)\,, 
\nn & a^B(k^B) =  Z_{a}^{\frac{1}{2}}\, a(k)\,, \quad
 k_{F}^B  =  k_F =\mu \, {\tilde{k}}_F \,,
\nonumber
\end{align}
\begin{align}
& Z_{\Psi}  = \frac{Z_2^{d-m+1} } { Z_0\, {Z_1 }^{d-m-1}}\,,\quad
Z_{a}  = \frac{Z_{3}\, Z_2^{d-m}} {Z_0\, {Z_1 }^{d-m-1}}\,,\nn
&  e^B=  Z_{e}\,e\,\mu^{\frac{x}{2}}\,, \quad
 Z_{e}= \frac{  Z_{4} \, Z_2^{\frac{d-m} {2} -1}} 
 {\sqrt{ Z_0\, Z_{3}} \, {Z_1 }^{\frac{d-m-1} {2}} }\,.
\end{align}
Here,
\begin{align}
Z_{\zeta}  =  1 + A_{\zeta}\,.
\end{align}
The superscript ``B'' denotes the bare fields, couplings, and momenta.
In Eq.~(\ref{act7}), there is a freedom to change 
the renormalizations of the fields 
and the renormalization of momentum
without affecting the action.
Here we fix the freedom by requiring 
that  $\delta_{k^B} = \delta_k$.
This amounts to measuring scaling dimensions 
of all other quantities relative to that of $\delta_k$.

\begin{figure}[]
\begin{center}
\subfigure[]{
\includegraphics[scale=0.2]{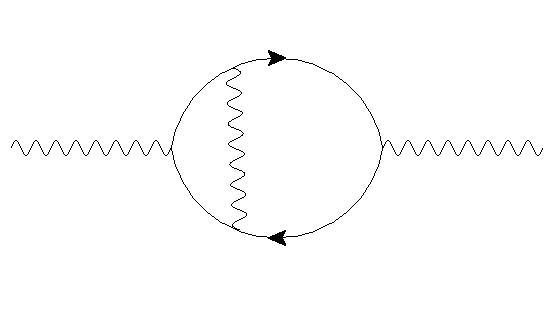} 
\label{fig:bos2a}
}\\
\subfigure[]{
\includegraphics[scale=0.2]{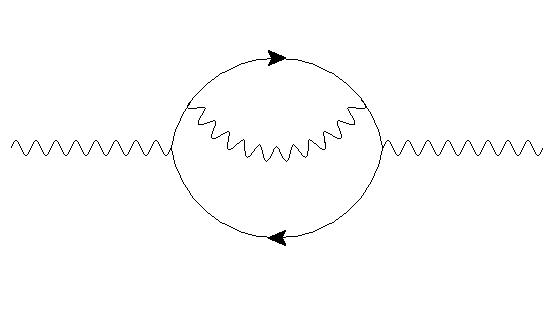} 
\label{fig:bos2b}
}
\subfigure[]{
\includegraphics[scale=0.2]{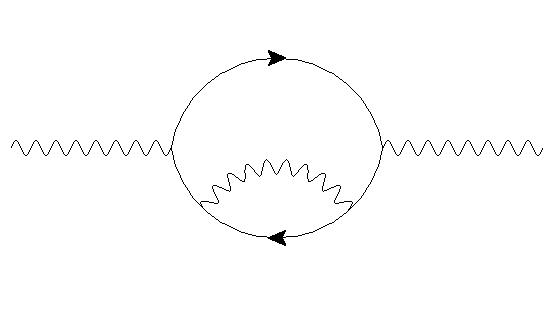} 
\label{fig:bos2c}
}\\
\subfigure[]{
\includegraphics[scale=0.2]{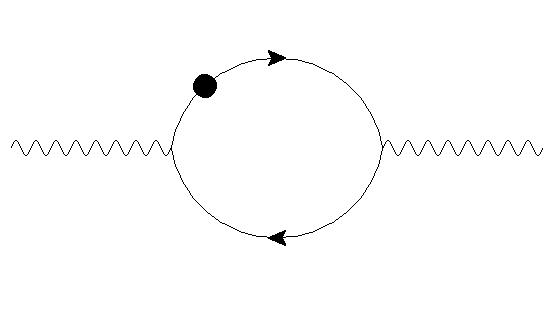} 
\label{fig:bos2d}}
\subfigure[]{
\includegraphics[scale=0.2]{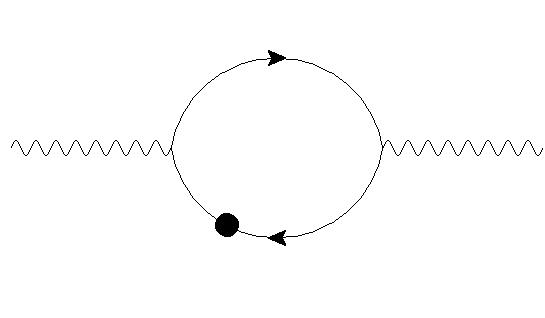} 
\label{fig:bos2e}
}
\end{center}
\caption{The two-loop diagrams contributing to the boson self-energy.
Each black dot denotes the one-loop counterterm for the fermion self-energy.
}
\label{fig:2loopbos}
\end{figure}

Let $z$ be the dynamical critical exponent, $\tilde z$ be the critical exponent along the extra spatial dimensions, $\beta_e $ be the beta function for the coupling $e$, $ \beta_{k_F}$ be the beta function for ${\tilde k}_F$,
and $\eta_\Psi$ ($\eta_\phi$) be
the anomalous dimension for the fermions (gauge boson). These are explicitly given by:
\begin{align}
& z = 1 + \frac{ \partial \ln (Z_0 / Z_2 ) }{\partial \ln \mu}\, , \quad
\tilde{z} = 1 + \frac{ \partial \ln (Z_1 / Z_2) }{\partial \ln \mu}\, , \nn
&
\eta_\Psi =  \frac{1}{2} \frac{ \partial \ln Z_\Psi}{\partial \ln \mu} \, , 
\quad \eta_{a} = \frac{1}{2} \frac{ \partial \ln Z_{a}}{\partial \ln \mu} \,,\nn
&
\beta_{k_F}({\tilde k}_F) =   \frac{\partial {\tilde k}_F}{\partial \ln \mu} \, , \quad
\beta_{e} =  \frac{\partial e}{\partial \ln \mu}\, .
\label{betaa1}
\end{align}
In the $\epsilon \rightarrow 0$ limit, we require solutions of the form:
\begin{align}
\label{solexp}
& z=z^{(0)}\,,\quad \tilde z={\tilde z}^{(0)}\,,\quad
\eta_\Psi = \eta_\Psi^{(0)}   +  \eta_\Psi^{(1)} \, \epsilon \,,\nn
&
 \eta_a = \eta_a^{(0)}   +  \eta_a^{(1)} \, \epsilon \,,\quad
\beta_e = \beta_e^{(0)}  + \beta_e ^{(1)} \, \epsilon \,.
\end{align}

\subsection{RG flows at one-loop order}

To one-loop order, the counterterms are given by $Z_\zeta = 1 + \frac{Z_{\zeta}^{(1)}} {\epsilon}\,.$ Collecting all the results, 
we find that only
\begin{align}
Z_{0}^{(1)}  & =  -\frac{  u_0\, \tilde{e}} {N}\,, \quad
Z_{1}^{(1)}  = - \frac{  u_1 \, \tilde{e} } {N} \,, \text{ and }
Z_{4}^{(1)}   = -\frac{  u_4 \, \tilde{e} } {N}
\end{align}
are nonzero, where
\begin{align}
\tilde{e}  =\frac{e^{ \frac{2 \,(m+1) } {3} }}
{  {\tilde{k}}_F ^{ \frac{(m-1) (2-m)}{6}   }}\,.
\end{align}

Then the one-loop beta functions, that dictate the flow of $\tilde k_F$ and $e$
with the increasing energy scale $\mu$, are given by:
\begin{align}
&  \beta_{k_F} =  - {\tilde k}_F \,,
\quad (1-z)\, Z_0= -\beta_{e}  \,   \frac{\partial Z_0} {\partial e}
  + {\tilde k}_F \,  \frac{\partial Z_0} {\partial \tilde{k}_F}   \,, \nn
& (1-\tilde z)\, Z_1= - \beta_{e} \,   \frac{\partial Z_1} {\partial e}
 + {\tilde k}_F \,  \frac{\partial Z_1} {\partial \tilde{k}_F}   \,, \nn
& \frac{ \beta_{e} } {e}
 = - \frac{  \epsilon} {2}
+
\frac{1}{2} \left [\frac{ \left (2-m \right ) \tilde{z}}{m+1} 
+ z-2  +\frac{m}{2}
\right ]  \,,\nn
&  \eta_\Psi = \eta_{a} 
=  -\frac{\epsilon }{2}  \left( 1 -\tilde{z}\right)
- \frac{z}{2}
+ \frac{ 3 - (2-m) \,\tilde{z} } {2 \,(m+1)}
\,.
\label{beta10}
\end{align}

\begin{figure}[]
\begin{center}
\subfigure[]{
\includegraphics[scale=0.2]{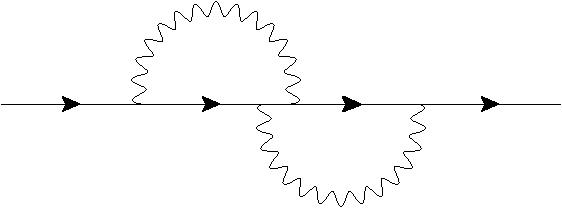} 
\label{fig:fer2a}
}\\
\subfigure[]{
\includegraphics[scale=0.2]{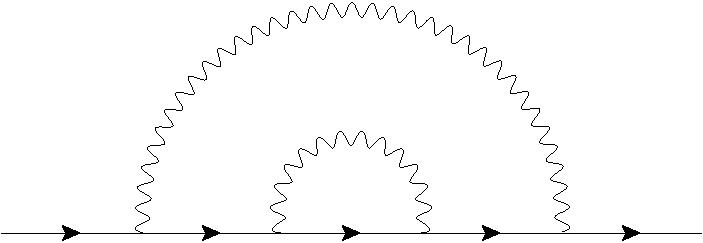} 
\label{fig:fer2b}
}
\subfigure[]{
\includegraphics[scale=0.2]{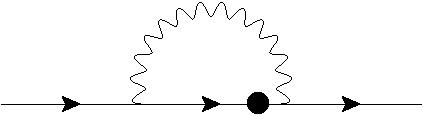} 
\label{fig:fer2c}
}
\end{center}
\caption{The two-loop diagrams contributing to the fermion self-energy.
Each black dot denotes the one-loop counterterm for the fermion self-energy.
}
\label{fig:2loopferm}
\end{figure}

Solving these equations using the required form outlined in Eq.~(\ref{solexp}), we get:
\begin{align}
z &= 1 + \frac{  (m+1) \,u_0 \,\tilde e }   
{  3 \,N  -  (m+1)\, u_1\,\tilde e}\,,
\, \,
 \tilde z  =  1 + \frac{  (m+1) \,u_1 \,\tilde e }   
 { 3 \,N  -(m+1)\, u_1\,\tilde e}\,,
 \nn \frac{\beta_e}{e}   & = 
 -\frac{\epsilon }{2} - \frac{ (m-1) (2-m)} {4\, (m+1)} 
\nn
& \qquad + \frac{ (m+1) \,u_0 +(2-m) \,u_1 -  2\, (m+1)\, u_4} 
{6 \, N} \,\tilde{e} \,.
\end{align}
The first term indicates that $e$ 
remains strictly relevant in the infrared (IR) at $d=d_c(m)$ for $1< m < 2\,.$
However, the second term implies that the higher order corrections are controlled not by $e$,
but by an effective coupling $\tilde e\,.$
Indeed, the scaling dimension of $\tilde e$ vanishes at $d_c $ for $1\leq m \leq 2\,.$
The beta function of this effective coupling is given by:
\begin{align}
& - \frac{\beta_{\tilde e} }  {\tilde e}
 \nn &
 = \frac{(m+1) \,\epsilon}{3} 
- \frac{ \left (m+1 \right ) 
\left[\, \left (m+1 \right ) \left (  u_0  - 2 \,u_4 \right )
+ \left ( 2-m\right ) u_ 1 \, \right ]}
{9 \, N} \,{\tilde e}\,.
\end{align}
The interacting fixed point is obtained from $\beta_{\tilde e} = 0$, and takes the form:
\begin{align}
{\tilde{e}}^*   = \frac{3 \,N\, \epsilon }
{  \left (m+1 \right ) \left( u_0 - 2\,u_4 \right)
+(2-m)\, u_1 } 
+\mathcal{O} \left( \epsilon^2 \right) .
\end{align}
It can be checked that this is an IR stable fixed point by computing the first derivative of $\beta_{\tilde e}\,.$
The critical exponents at this stable fixed point are given by:
\begin{align}
\label{critex}
& z^*= 1+\frac{(m+1) \,u_0 \,\epsilon }
{\left (m+1 \right ) \left( u_0 - 2\,u_4 \right)  
+ (2-m) \,u_1}\,,\nn
& {\tilde z}^*= 1+\frac{(m+1) \,u_1 \,\epsilon }
{ \left (m+1 \right ) \left( u_0 - 2\,u_4 \right)
+ (2-m) \,u_1}
 \,,\nn
& \eta_\Psi^* = \eta_a^* = - \frac{(m+1) \,u_0 + (2-m) \,u_1}
{ (m+1) \left(u_0 - 2 \, u_4 \right)+ (2-m) \, u_1}
\, \frac{\epsilon}{2} \,.
\end{align}

\subsection{Higher-loop corrections}


\begin{figure}[t]
\begin{center}
\includegraphics[scale=0.5]{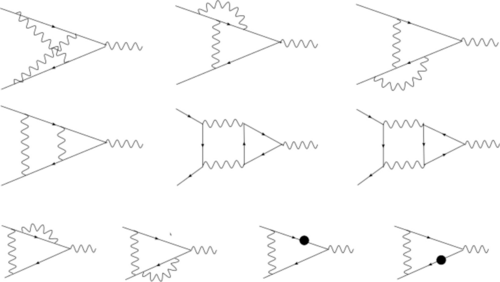} 
\end{center}
\caption{The two-loop diagrams contributing to the vertex correction.
Each black dot denotes the one-loop counterterm for the fermion self-energy.
}
\label{fig:2loopvert}
\end{figure}

We will now discuss the implications of the higher-loop corrections, without actually computing the Feynman diagrams. For $m>1 $, we expect a nontrivial UV/IR mixing to be present, as was found in Ref.~\cite{ips-uv-ir1,ips-uv-ir2}, which makes the results one-loop exact. In other words, all higher-loop
corrections would vanish for $m>1$ in the limit $k_F \rightarrow 0\,,$ due to suppression of the results by
positive powers of $k_F\,.$
For $m=1$, we will use the arguments and results of Ref.~\cite{Lee-Dalid} to assume a generic form of the corrections coming from two-loop diagrams. Henceforth, we will just focus on $m=1$ in this subsection.

The two-loop diagrams for the boson self-energy have been shown in Fig.~\ref{fig:2loopbos}.
The value should turn out to be UV finite, and hence will renormalize the factor $\beta(\frac{5}{2},1)$ (see Eq.~\ref{eqbetad}) by a finite amount $\beta_2 = \frac{\kappa\,  {\tilde e}} {N}\,,$ where $\kappa$ is a finite number. Then the bosonic propagator at this order will take the form:
\begin{align}
\label{babos2}
D_2 (q) 
= \frac{1}{{\vec{L}}_{(q)}^2 
+ \frac{\left[  \beta \big(\frac{5} {2},2 \big) + \frac{\kappa\, \tilde e } 
{N}\right]  e^2  \mu^\epsilon   } 
{   |\vec{L}_{(q)}|}  
\times \frac{  k_0^2 + \left ( \epsilon- \frac{1}{2} \right)\,{\tilde {\vec K}}^2  }
{|\vec K|^{ \frac{1 }  {2}+\epsilon }}\,.
} \,.  
\end{align}
From this, the fermion self-energy now receives a correction
\begin{align}
\Sigma_{2}^{(1)}(k) =& \left [
\left \lbrace \frac{ \beta \big(\frac{5} {2},2 \big)} 
{\beta \big(\frac{5} {2},2 \big) +\frac{  \kappa \, \tilde e}
{N} }\right \rbrace ^{\frac{1}{3}} 
-1 \right ] \Sigma_1(k) \nn
= & -\frac { \kappa \, \tilde e } 
{3 \,N\, \beta \big(\frac{5} {2},2 \big) }\,\Sigma_1(k)
+ \text{ finite terms} \,.
\end{align}

Now the two-loop fermion self-energy diagrams (see Fig.~\ref{fig:2loopferm}), after taking into account the counterterms obtained from one-loop
corrections, take the form:
\begin{align}
 \Sigma_{2}^{(2)}(k) =&
-\frac{ \mathrm{i} \,{\tilde e}^2 
\Big[\,
 \tilde v_{0} \,\gamma_0\,q_0 +
\tilde v_1 \, \left( \tilde{\vec \Gamma } \cdot \tilde{\vec Q } \right) 
 + w\, \gamma_{d-1}\,\delta_k \,\Big ]    }
{ N^2\,\epsilon} 
\nn & + \text{ finite terms}\,.
\end{align}
Adding the two, generically the total two-loop fermion self-energy can be written as:
\begin{align}
 \Sigma_{2}^{tot}(k) =&
-\frac{ \mathrm{i} \, \tilde e^{2} 
\Big[\,v_{0} \,\gamma_0\,q_0 + v_1 \, \left( \tilde{\vec \Gamma } \cdot \tilde{\vec Q } \right) 
 + w\, \gamma_{d-1}\,\delta_k \,\Big ]    }
{ N^2\,\epsilon} 
\nn & + \text{ finite terms}\,,
\end{align}
where $v_0 = u_0 +\tilde{v}_0 $ and $v_1 = u_1 +\tilde{v}_1 $.

There will also be a divergent vertex correction (see Fig.~\ref{fig:2loopvert}) which will lead to a nonzero $Z_4^{(1)}$
of the form $-\frac{\tilde e^{2}  \,y}{N^2}\,.$
All these now lead to the nonzero coefficients:
\begin{align}
& Z_{0}^{(1)} =  
-\frac{  u_0\, \tilde{e}} {N}
-\frac{  v_0\, \tilde{e}^2  } {N^2 } \,,\quad
 Z_{1}^{(1)}  = 
 - \frac{  u_1 \, \tilde{e} } {N} 
-\frac{ v_1 \, \tilde{e}^2  } {N^2 } \,,\nn
 & Z_{2}^{(1)}  = 
-\frac{ w\, \tilde{e}^2  } {N^2 } \,,\quad
Z_{4}^{(1)}  =
-\frac{  u_4 \, \tilde{e} } {N}
-\frac{  y\,\tilde{e}^2  } {N^2 }\,,
\end{align}
resulting in
\begin{widetext}
\begin{align}
\frac{ \beta_{\tilde e}}  {\tilde e} 
=  
-\frac{2  \left(2 \,u_1 \tilde{e} + 3 \,N\right)  \epsilon} {9 \,N}
-
\frac{2 \left(2 \,u_0+u_1-4 \, u_4\right) \tilde{e}}
{9 N}
+ \frac{4 \left[ 
-u_1^2-2 \,u_0 \,u_1 + 4 \,u_4 \,u_1
-3 \left(2 \, v_0 +  v_1 -3\, w \right)
+ 12 \,y \right ]
 \tilde{e}^2 }
{27\, N^2}  \,.
\end{align}
\end{widetext}
At the fixed point, we now have:
\begin{align}
 \frac{\tilde e^*} {N} 
& = 
\frac{3 \,\epsilon }{2\, u_0 + u_1-4 \,u_4}
-\frac{18
 \left(2 \,v_0+ v_1-3 \,w-4 \,y\right)
 \epsilon ^2 }
{\left(2 \,u_0+ u_1-4 \,u_4\right)^3} 
\nn & \qquad+
\mathcal{O} \big(\epsilon^3 \big) \,.
\end{align}
This shows that the nature of the stable non-Fermi liquid fixed point remains unchanged, although
its location (as well as any critical scaling) gets corrected by one higher power of $\epsilon$.

\subsection{Renormalization of the $2k_F$ scattering amplitude}
\label{scatter2kf}

\begin{figure}[]
\begin{center}
\subfigure[]{
\includegraphics[scale=0.25]{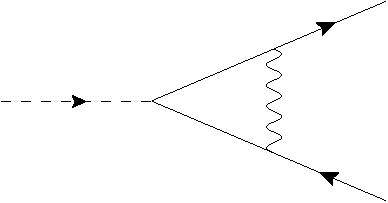} 
\label{fig:2kf1}
}\quad
\subfigure[]{
\includegraphics[scale=0.25]{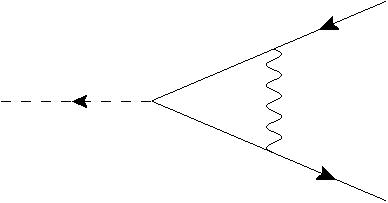} 
\label{fig:2kf2}
}
\end{center}
\caption{The one-loop diagrams contributing to the $2k_F$ scattering amplitude.
}
\label{fig:2kf}
\end{figure}

In order to examine how the back-scattering is affected by the interactions with the gauge bosons in the non-Fermi liquid state, we consider an operator which carries 
momentum $2 k_F$ as follows:
\begin{align}
\label{scat2k1}
& S_{2k_F} \nn & =  -2\, g_{2k_F} \, \mu
 \sum_j \int dk \left[ 
(\psi_{+,j}^\dagger (k) \psi_{-,j} (k) + \psi_{-,j}^\dagger (k) \psi_{+,j} (k) \right]\nn
& =  \mathrm{i}\, g_{2k_F} \, \mu \int dk \left[ 
\Psi^T (k) \gamma_0 \Psi (-k) + \bar{\Psi} (k) \gamma_0 \bar{\Psi}^T (-k) \right] ,
\end{align}
where $g_{2k_F}$ is the source. 
To cancel UV divergences, 
we need to add a counterterm of the form:
\begin{align}
 S_{2k_F}^{CT}  
& =  \mathrm{i}\,
 g_{2k_F} \, \mu  \left (Z_{2k_F} - 1 \right )
  \int dk
\Big[  \Psi^T (k) \,\gamma_0 \Psi (-k) 
\nn 
& \hspace{ 4.2 cm } + 
\bar{\Psi} (k) \,\gamma_0 \bar{\Psi}^T (-k) \Big ] \, ,
\end{align}
which renormalizes the insertion into
\begin{align}
 S_{2k_F}^{ren} 
& =  \mathrm{i}\, g_{2k_F}^B
\int dk^B \Big[  
\left( \Psi^B (k) \right )^T  \gamma_0 \,\Psi^B (-k) 
\nn & \hspace{ 2.75 cm}
+ {\bar{\Psi}}^B  (k)\, \gamma_0 \,
\left( {\bar{\Psi}}^B (-k) \right) ^T 
 \Big ] \, ,
\nn & g_{2k_F}^B=  Z_{g}\,g_{2k_F}\,\mu \,, \quad
 Z_{2k_F}=  Z_g\,Z_2  \,.
\end{align}
Here,
 \begin{align}
 Z_{2k_F} =  1+ \frac{Z_{2k_F}^{(1)}} {\epsilon}
\end{align}
to one-loop order.
The loop calculations involving the diagrams in Fig.~\ref{fig:2kf} have been shown in details in Appendix~\ref{oneloop2kf}, which lead to:
\begin{align}
& Z_{2k_F}^{(1)} =  \begin{cases}
- \frac{  0.0774559 \, \tilde e }  {N}   
& \text{ for } m=1 \\
0
& \text{ for } m=2
\end{cases} \,,
\end{align}
This gives the beta function for $ g_{2k_F}$ as:
\begin{align}
\beta_g  =  - g_{2k_F} \left( 1 - \eta_g \right),
\end{align}
with anomalous dimension $\eta_g =- \frac{2 \,\tilde e\, u_g }{3 N} $, where $u_g= 0.0774559$ for $m=1$. The negative value of $\eta_g$ shows that the $2 k_F$ scattering amplitude is enhanced by fluctuations of the transverse gauge field. 
This is in contrast with the behaviour computed in the case of  in  the  Ising-nematic
quantum criticality, where the $2\,k_F$ scattering amplitude is suppressed \cite{Lee-Dalid} in the presence of the Ising-nematic critical bosons in $d_{\text{phys}}=2$.
Note that this scattering amplitude can also be interpreted as an instability in the charge density wave (CDW) channel, which therefore (due to its negative anomalous dimension) turns out to be a serious competitor for the transverse gauge field  criticality for $m=1$.

\subsection{Thermodynamic quantities }

The scaling of thermodynamic quantities are different 
from observables which are local in momentum space.
This is because all low energy modes near the Fermi surface contribute
to the thermodynamic responses. Here, we will outline the expectations of a general scaling analysis. 

In this paper, the momentum components $ k_{d-m}$ and $\mathbf L_{(k)}$ have scaling dimensions one, $k_0$ has scaling dimension $z$, and the $d_c(m) -m-\epsilon -1= \frac{2-m}{m+1} - \epsilon$ momentum components with Dirac dispersion have scaling dimension $\tilde z$.
Note that for Ising-nematic critical point $\tilde z = z$.

In order to examine the scaling behavior of thermodynamic quantities,
we consider the free energy density at finite temperature $T$.
In a system with $d_{\text{phys}}= m+1$ spatial dimensions, fermionic dynamical critical exponent $z$, and $ \frac{2-m}{m+1} - \epsilon $ auxiliary dimensions with critical exponent $\tilde z$, the free energy density $F(T)$ has the scaling dimension $ [F(T)] 
= d_{\text{phys}}+z + \left( \frac{2-m}{m+1} - \epsilon  \right) \tilde{z} $, if it were independent of any UV cut-off scale. However, when a critical boson couples with fermions on all parts of the Fermi surface, the entire Fermi surface becomes hot. As a result, we expect a hyperscaling violation, such that the the singular part of the free energy density depends on the size of the Fermi surface \cite{LEE2008,ips-subir}. The largest momentum along the $\mathbf L_{(k)}$ direction is set by the Fermi momentum $k_F$, and hence the free energy density should have the scaling form:
\begin{align}
F(T) & \sim   k_F^{m/2}\,
T^{1+ \frac{d_{\text{phys}}-m} {z} 
+ \frac{\left( \frac{2-m}{m+1} - \epsilon  \right) \tilde{z}} {z}} 
\nn
& \sim k_F^{m/2}\,
T^{1+ \frac{ 1+\left( \frac{2-m}{m+1} - \epsilon  \right) \tilde{z}} {z}}
\end{align}
in the presence of an $m$-dimensional Fermi surface, with an effective scaling dimension $ 
[F(T)]_{\text{eff}} =  1+  z + \left( \frac{2-m}{m+1} - \epsilon  \right) \tilde{z} $. 
From this scaling form, we can extract the temperature dependence of various observables within the quantum critical region. For example, the specific heat should scales as
$ C  \propto T^{ \frac{ 1+\left( \frac{2-m}{m+1} - \epsilon  \right) \tilde{z}} {z}}$.

The current operator is given by $J(T)=\frac{\delta F(T)}{\delta A}$, where $A $ is the vector potential with scaling dimension one. Hence, it should have the scaling form:
\begin{align}
J(T)   
\sim k_F^{m/2}\,\,
T^{ 1 +\frac{ \left( \frac{2-m}{m+1} - \epsilon  \right) \tilde{z}} {z}}\,,
\end{align}
with an effective scaling dimension $ 
[J(T)]_{\text{eff}} =   z + \left( \frac{2-m}{m+1} - \epsilon  \right) \tilde{z} $.
Then using the Kubo formula, we can infer that the effective scaling dimension of the optical conductivity is
\begin{align}
&[\sigma(\omega) ]_{\text{eff}}  
= 2\,[J(T)]_{\text{eff}}- z-[\text{volume in } k\text{-space}]_{\text{eff}} \nn
& = 2\,z +2\left( \frac{2-m}{m+1} - \epsilon  \right) \tilde{z}
- z- z- 1
-  \left( \frac{2-m}{m+1} - \epsilon  \right) \tilde{z} \nn
& = -1 +\left( \frac{2-m}{m+1} - \epsilon  \right) \tilde{z}\,,
\end{align}
leading to the scaling form:
\begin{align}
&[\sigma(\omega \gg T) ] 
\propto \omega^{-\frac{1}{z}+\frac{ \left( \frac{2-m}{m+1} - \epsilon  \right) \tilde{z}} {z}}\,,
\end{align}
where $\omega $ is the frequency of the applied AC electric field.


\section{Model involving two $U(1)$ transverse gauge fields}
\label{modelu2}

In this section, we consider the $m$-dimensional Fermi surfaces of two different kinds of fermions (denoted by subscripts $1$ and $2$) coupled to two U(1) gauge
fields, $a_c$ and $a_s$, in the context of deconfined Mott transition and deconfined metal-metal transition studied in Ref.~\cite{debanjan} (for $m=1$).
The theoretical motivation of Ref.~\cite{debanjan} was to study a distinct class of quantum phase transitions between a Fermi liquid and a Mott insulator \cite{sachdev_2011}, or between two metals that have Fermi surfaces with finite but different sizes on either side of the transition \cite{keimer,vojta}. These have been dubbed by the authors as deconfined Mott transition (DMT), and deconfined metal-metal transition (DM$^2$T), respectively. These problems can be formulated using a fictitious / emergent $U(2)$ gauge field, but the authors showed that this non-abelian gauge field is `quasi-abelianized' such that a related $U(1) \times U(1)$ gauge theory can capture many essential features.
In this $U(1) \times U(1)$ gauge theory, the fermion fields $\psi_{1,\pm,j}$ and $\psi_{2,\pm,j}$ carry negative charges under the
even ($a_c + a_s$) and odd ($a_c - a_s$) combinations of the gauge fields.
We revisit this problem using our dimensional regularization scheme because using this technique, we can study this system in generic dimensions, and also perform higher-loop diagrams giving order by order corrections in $\epsilon$.

The action takes the form:
\begin{widetext}
\begin{align}
S & = \sum \limits_{\alpha=1,2}  \sum \limits_{p=\pm} \sum_{j=1}^N \int dk\,
\psi_{\alpha,p,j}^\dagger (k)
\Bigl[ \mathrm{i}\, k_0   +  p  \,k_{d-m} +  {\vec L}_{(k)}^2  \Bigr] \psi_{\alpha,p,j}(k)
 + \frac{1}{2} \int  dk \, {\vec L}_{(k)}^2 \left[ 
  a_c^\dagger(k) \, a_c(k) +  a_s^\dagger(k) \, a_s(k) \right ] \nonumber \\
 & \quad
+\sum_{\alpha=1,2} \sum_{p=\pm} p\sum_{j=1}^N \int dk  \, dq
\left[
\frac{ (-1)^\alpha \, e_s}{\sqrt{N}}  \,  a_s(q) \,  \psi^\dagger_{\alpha,p,j}(k+q) 
\, \psi_{\alpha, p,j}(k) 
-  \frac{e_c}{\sqrt{N}} \, a_c(q) \,  \psi^\dagger_{\alpha,p,j}(k+q) 
\, \psi_{\alpha, p,j}(k)\right ] ,
\label{actu2}
\end{align}
\end{widetext}
where $e_c $ and $e_s $ denote the gauge couplings for the gauge fields $a_c $ and $a_s $ respectively.
We will perform dimensional regularization on this action and determine the RG fixed points.
Our formalism allows us to extend the discussion beyond $m=1$, and also to easily compute higher-loop corrections.

\subsection{Dimensional regularization}

Proceeding as in the single transverse gauge field case, we add artificial co-dimensions for dimensional regularization after introducing the two-component spinors:
\begin{align}
& \Psi_{\alpha,j}^T(k) = \left( 
\psi_{\alpha,+,j}(k),
\psi_{\alpha,-,j}^\dagger(-k)
\right) \text{ and } \bar \Psi_{\alpha,j} \equiv \Psi_{\alpha, j}^\dagger \,\gamma_0\,,
\nn
& \text{with } \alpha =1,2 \,.
\end{align}  
The dressed gauge boson propagators include the one-loop self-energies given by:
\begin{align}
\label{babos2}
& \Pi^c_1 (k)  =
-\frac{ \beta(d,m)\,  e_c^2 \, \mu^x \left(  \mu \, {\tilde{k}}_F  \right )^{\frac{m-1}{2}} 
 } 
{   |\vec{L}_{(q)}|}  
 \nn & \hspace{ 1.5 cm} \times
 \left[ k_0^2 + ( m+1-d)\,{\tilde {\vec K}}^2 \right] 
|\vec K|^{d-m-2} \,, 
\end{align}
and
\begin{align}
& \Pi^s_1 (k)  =
-\frac{ \beta(d,m)\,  e_s^2 \, \mu^x \left(  \mu \, {\tilde{k}}_F  \right )^{\frac{m-1}{2}} 
 } 
{   |\vec{L}_{(q)}|}  
 \nn & \hspace{ 1.5 cm} \times
 \left[ k_0^2 + ( m+1-d)\,{\tilde {\vec K}}^2 \right] 
|\vec K|^{d-m-2} \,,
\end{align}
for the $a_c$ and $a_s$ gauge fields, respectively.
This implies that the one-loop fermion self-energy for both $\Psi_{1,j}$ and $\Psi_{2,j}$ now takes the form:
\begin{align}
\label{sigmau1}
\Sigma_1(q) = &
-\frac{ \mathrm{i} \left( e_c^{\frac{2\,(m+1)} {3} } + e_s^{\frac{2\,(m+1)} {3} } \, 
\right)   }
{ N \, {\tilde{k}}_F ^{ \frac{(m-1)(2-m) } {6}}}
\frac{u_0\,\gamma_0\,q_0 + u_1  \left( \tilde{\vec \Gamma } \cdot \tilde{\vec Q } \right) }
{\epsilon} 
\nn &  + \text{ finite terms} \,,
\end{align}
with the critical dimension $d_c = \left(  m+\frac{3}{m+1}\right)$, $u_0$ and $u_1$ (See Eq.~\ref{valu}) having the same values as for the $U(1)$ case.

The counterterms take the same form 
as the original local action:
\begin{widetext}
\begin{align}
S_{CT}  = &  \sum_{\alpha,j} \int dk \, \bar \Psi_{\alpha,j}(k)
\, \mathrm{i} \,\Bigl[ 
A_{0} \,  \gamma_0 \,k_0 + A_{1} \,\tilde{\vec \Gamma} \cdot \tilde{ \vec K} 
+   A_2 \, \gamma_{d-m} \, \delta_k 
 \Bigr] \Psi_{\alpha,j}(k) \,  \exp \Big \lbrace \frac {{\vec{L}}_{(k)}^2}  { \mu \, {\tilde{k}}_F } \Big \rbrace
 + \frac{A_{3_s}}{2} \int  dk\,
 {\vec{L}}_{(k)}^2\,   a_s^\dagger(k) \, a_s(k)
\nn &+  \frac{A_{3_c}}{2} \int  dk \,
 {\vec{L}}_{(k)}^2\,  a_c^\dagger(k) \, a_c(k)  
-  A_{4_c} \frac{  e_c \, \mu^{x/2} }{\sqrt{N}} \sum_{\alpha,j}  
\int dk \, dq  \,
a_c(q) \,  \bar \Psi_{\alpha, j}(k+q) \,\gamma_{0} \, \Psi_{\alpha,j}(k) 
\nn & +  A_{4_s} \frac{ e_s \, \mu^{x/2} }{\sqrt{N}} \sum_{\alpha,j}  (-1)^\alpha
\int \frac{d^{d+1}k \, d^{d+1}q}{(2\pi)^{2d+2}}  \,
a_s (q) \,  \bar \Psi_{\alpha, j}(k+q) \,\gamma_{0} \, \Psi_{\alpha,j}(k) \, ,
\end{align}
where 
\begin{align}
A_{\zeta} = 
\sum_{\lambda=1}^\infty \frac{Z^{(\lambda)}_{ \zeta}
(e_,\tilde{k}_F)}{\epsilon^\lambda}  \text{  with }  \zeta=0,1,2,3_c,3_s , 4_c,4_s\,.
\end{align}
\end{widetext}
We have taken into account the exchange symmetry: $ \Psi_{1,j} \leftrightarrow \Psi_{2,j}\,,\,\,
a_s \rightarrow - a_s \,,$ which was assumed in Ref.~\cite{debanjan}, and here it means that
both $\Psi_{1,j}$ and $\Psi_{2,j}$ have the same wavefunction renormalization $ Z_{\Psi}^{1/2}$.

Adding the counterterms to the original action,
we obtain the renormalized action:
\begin{widetext}
\begin{align}
\label{act8}
S_{ren}  = & \sum_{\alpha,j} \int dk^B  \bar \Psi^B_{\alpha,j}(k^B)
\, \mathrm{i} \left[   \gamma_0 \,k_0^B + \tilde{\vec \Gamma} \cdot \tilde{ \vec K}^B 
+    \gamma_{d-m} \, \delta_k \right ] \Psi^B_{\alpha,j}(k^B)
 \,  \exp \Big \lbrace \frac {{\vec{L}}_{(k^B)}^2}  { \mu \, {\tilde{k}}_F^B } \Big \rbrace
 + \frac{1}{2} \int  dk^B\,
 {\vec{L}}_{(k^B)}^2\,  {a^B_c }^\dagger(k^B) \, \,\,a^B_c(k^B)
\nn &+  \frac{1}{2} \int  dk^B \,
 {\vec{L}}_{(k^B)}^2\, { a^B_s }^\dagger(k^B) \,\,\, a_s^B(k^B)  
-   \frac{ e_c^B  }{\sqrt{N}} \sum_{\alpha,j}  
\int dk^B \, dq^B  \,
a^B_c (q^B) \,  \bar \Psi^B_{\alpha, j}(k^B+q^B) \,\gamma_{0} \, \Psi^B_{\alpha,j}(k^B) 
\nn & +    \frac{ e_s^B  }
{\sqrt{N}} \sum_{\alpha,j}  (-1)^\alpha
\int dk^B \, dq^B  \,
a_s^B (q^B) \,  \bar \Psi^B_{\alpha, j}(k^B+q^B) \,\gamma_{0} \, \Psi^B_{\alpha,j}(k^B)  \, ,
\end{align}
remembering that $\delta_{k^B} =\delta_k\,.$
Here
\begin{align}
& k_{0}^B = \frac{Z_0} {Z_2}\,k_0\,,\quad
\tilde{\vec K}^B =   \frac{Z_1} {Z_2} \, \tilde{\vec K} \, , \quad
k_{d-m}^B =  k_{d-m} \, , 
\quad {\vec{L}}_{(k^B)}  =  {\vec{L}}_{(k)} \,, \quad
k_{F}^B  =  k_F =\mu \, {\tilde{k}}_F \,, \quad
\Psi_j^B(k^B)  =   Z_{\Psi}^{\frac{1}{2}}\, \Psi_j(k)\,,
\nn &
 a_c^B(k^B) =  Z_{a_c}^{\frac{1}{2}}\, a_c(k)\,, \quad
 a_s^B(k^B) =  Z_{a_s}^{\frac{1}{2}}\, a_s(k)\,, \quad
  Z_{\Psi}  = \frac{Z_2^{d-m+1} } { Z_0\, {Z_1 }^{d-m-1}}\,,\quad
 Z_{a_c}  = \frac{Z_{3_s}\, Z_2^{d-m}} {Z_0\, {Z_1 }^{d-m-1}}\,,\quad
Z_{a_s}  = \frac{Z_{3_c}\, Z_2^{d-m}} {Z_0\, {Z_1 }^{d-m-1}}\,,\nn &
 e_c^B=  Z_{e_c}\,e_c\,\mu^{\frac{x}{2}}\,, \quad
 Z_{e_c}= \frac{  Z_{4} \, Z_2^{\frac{d-m} {2} -1}} 
 {\sqrt{ Z_0\, Z_{3_c}} \, {Z_1 }^{\frac{d-m-1} {2}} }\,,\quad
  e_s^B=  Z_{e_s}\,e_s\,\mu^{\frac{x}{2}}\,, \quad
 Z_{e_s}= \frac{  Z_{4} \, Z_2^{\frac{d-m} {2} -1}} 
 {\sqrt{ Z_0\, Z_{3_s}} \, {Z_1 }^{\frac{d-m-1} {2}} }\,, 
\end{align}
\end{widetext}
and
\begin{align}
Z_{\zeta}  =  1 + A_{\zeta}\,.
\end{align}
As before, the superscript ``B'' denotes the bare fields, couplings, and momenta.

As before, we will use the same notations, namely, $z$ for the dynamical critical exponent, $\tilde z$ for the critical exponent along the extra spatial dimensions, $ \beta_{k_F}$ for the beta function for ${\tilde k}_F$,
and $\eta_\psi$ for the anomalous dimension of the fermions. Since we have two gauge fields now, we will use
the symbols $\beta_{e_c} $ and $\beta_{e_s}$ to denote the beta functions for the couplings $e_c$ and $e_s$ respectively, which are explicitly given by:
\begin{align}
\beta_{e_c} =  \frac{\partial e_c}{\partial \ln \mu}\,,\quad
\beta_{e_s} =  \frac{\partial e_s}{\partial \ln \mu}\, .
\label{beta2}
\end{align}
The anomalous dimensions of these two bosons are indicated by:
\begin{align}
&\eta_{a_c} = \frac{1}{2} \frac{ \partial \ln Z_{a_c}}{\partial \ln \mu} \,,\quad
\eta_{a_s} = \frac{1}{2} \frac{ \partial \ln Z_{a_s}}{\partial \ln \mu} \,.
\end{align}

\subsection{RG flows at one-loop order}

To one-loop order, the counterterms are given by $Z_\zeta = 
1 + \frac{Z_{\zeta}^{(1)}} {\epsilon}\,.$
Here, 
\begin{align}
Z_{0}^{(1)}  &=    - \frac{ u_0 \left( \tilde{e}_c +{\tilde e}_s \right)  }
 {N } \,,\quad
 Z_{1}^{(1)} =   -\frac{ u_1 \left( \tilde{e}_c +{\tilde e}_s \right) }
 {N } \,,\nn
 \text{and }
Z_{4}^{(1)} &=   -\frac{ u_4 \left( \tilde{e}_c +{\tilde e}_s \right) }
 {N } 
\end{align}
are nonzero, where
\begin{align}
\tilde{e}_c  =\frac{e_c^{ \frac{2 \,(m+1) } {3} }}
{  {\tilde{k}}_F ^{ \frac{(m-1) (2-m)}{6}   }}
\text{ and }
\tilde{e}_s  =\frac{e_s^{ \frac{2 \,(m+1) } {3} }}
{  {\tilde{k}}_F ^{ \frac{(m-1) (2-m)}{6}   }}\,.
\end{align}

The one-loop beta functions are now given by:
\begin{align}
&  \beta_{k_F} =  - {\tilde k}_F \,,
\nn & (1-z)\, Z_0= -\beta_{e_c}  \,   \frac{\partial Z_0} {\partial e_c} 
-\beta_{e_s}  \,   \frac{\partial Z_0} {\partial e_s }
  + {\tilde k}_F \,  \frac{\partial Z_0} {\partial \tilde{k}_F}   \,, \nn
& (1-\tilde z)\, Z_1= - \beta_{e_c}  \,   \frac{\partial Z_1} {\partial e_c} 
-\beta_{e_s }  \,   \frac{\partial Z_1} {\partial e_s}
 + {\tilde k}_F \,  \frac{\partial Z_1} {\partial \tilde{k}_F}   \,, \nn
& \frac{ \beta_{e_c} } {e_c}
 = - \frac{  \epsilon} {2}
+
\frac{1}{2} \left [\frac{ \left (2-m \right ) \tilde{z}}{m+1} 
+ z-2  +\frac{m}{2}
\right ]   \,,\nn
& \frac{ \beta_{e_s} } {e_s}
 = - \frac{  \epsilon} {2}
+
\frac{1}{2} \left [\frac{ \left (2-m \right ) \tilde{z}}{m+1} 
+ z-2  +\frac{m}{2}
\right ]   \, .
\label{beta11}
\end{align}
Solving these equations, we get:
\begin{widetext}
\begin{align}
 -\frac{\beta_{e_c}} {e_c} =-\frac{\beta_{e_s}} {e_s}   
= 
\frac{\epsilon }{2} + \frac{ (m-1) (2-m)}{4 (m+1)} 
- \frac{ (m+1) \,u_0 +(2-m) \,u_1 -  2\, (m+1)\, u_4} 
{6 \, N} \left( \tilde{e}_c  + \tilde e_s \right) .
\end{align}
\end{widetext}
Again, it is clear that for generic $m$, the order by order loop corrections are controlled not by $e_c$ and $e_s $, but by the effective couplings $\tilde{e}_c $ and $ \tilde{e}_s \,.$
Hence we need to compute the RG flows from the beta functions of these effective couplings, which are given by:
\begin{widetext}
\begin{align}
-\frac{\beta_{\tilde e_c} }  {\tilde e_c} =-\frac{\beta_{\tilde e_s} }  {\tilde e_s} 
=
 \frac{(m+1) \,\epsilon}{3} 
 - \frac{ \left (m+1 \right ) 
\left[\, \left (m+1 \right ) \left (  u_0  - 2 \,u_4 \right )
+ \left ( 2-m\right ) u_ 1 \, \right ]
\left( \tilde{e}_c + \tilde e_s \right )
}
{9 \, N}\, .
\end{align}
\end{widetext}
The interacting fixed points are determined from the zeros of the above beta functions, and take the form:
\begin{align}
{\tilde{e}_c}^*  + {\tilde e}_s^*   = 
 \frac{3 \,N\, \epsilon }
{  \left (m+1 \right ) \left( u_0 - 2\,u_4 \right)
+(2-m)\, u_1 } 
+\mathcal{O} \left( \epsilon^2 \right) ,
\end{align}
which actually give rise to a fixed line, as found in Ref.~\cite{debanjan} for the case of $m=1\,.$
It can be checked that this is IR stable by computing the first derivative of the beta functions.
Hence, we have proven that the fixed line feature survives for critical Fermi surfaces of dimensions more than one.
The critical exponents at this stable fixed line take the same forms as in Eq.~(\ref{critex}).

\subsection{Higher-loop corrections}

\begin{figure*}[]
\begin{center}
\subfigure[]{
\includegraphics[width=0.22 \textwidth]{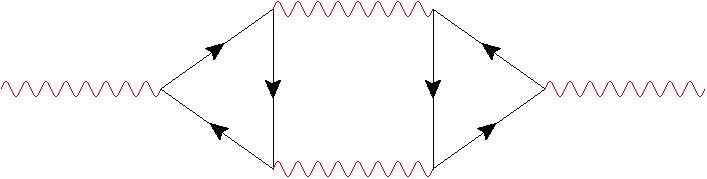} 
}
\subfigure[]{
\includegraphics[width=0.22 \textwidth]{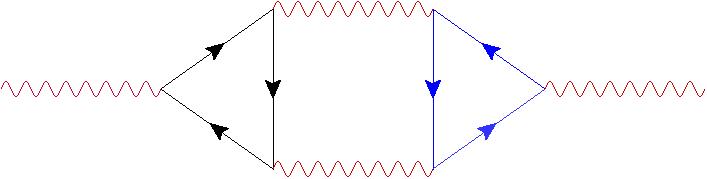} 
}
\subfigure[]{
\includegraphics[width=0.22 \textwidth]{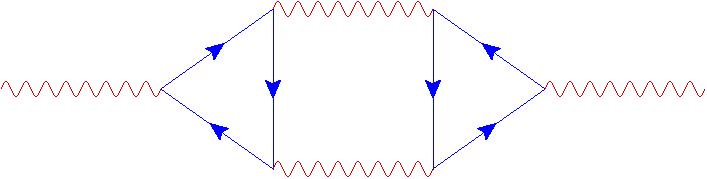} 
}
\subfigure[]{
\includegraphics[width=0.22 \textwidth]{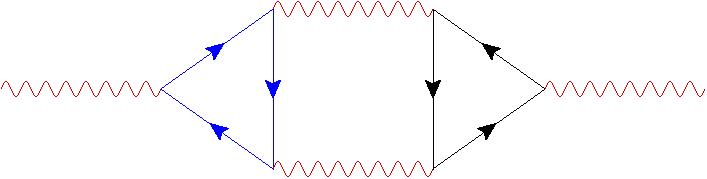} 
}
\subfigure[]{
\includegraphics[width=0.22 \textwidth]{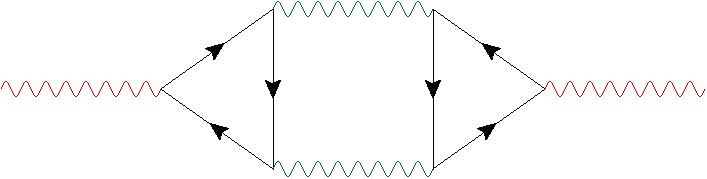} 
}
\subfigure[]{
\includegraphics[width=0.22 \textwidth]{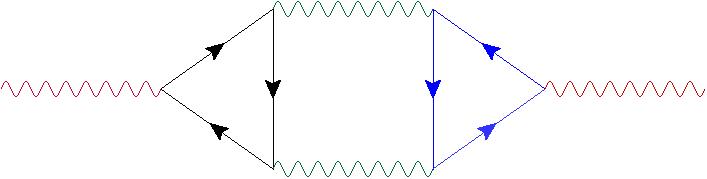} 
}
\subfigure[]{
\includegraphics[width=0.22 \textwidth]{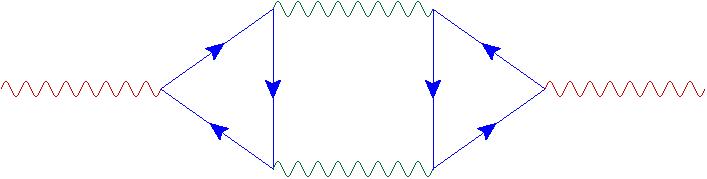} 
}
\subfigure[]{
\includegraphics[width=0.22 \textwidth]{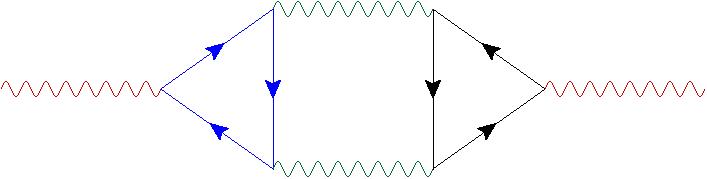} 
}
\subfigure[]{
\includegraphics[width=0.22 \textwidth]{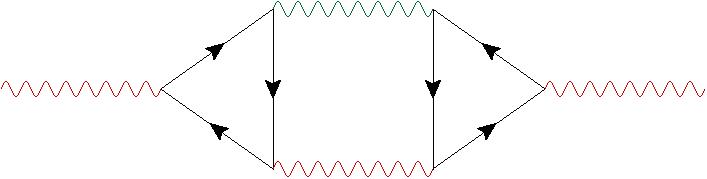} 
}
\subfigure[]{
\includegraphics[width=0.22 \textwidth]{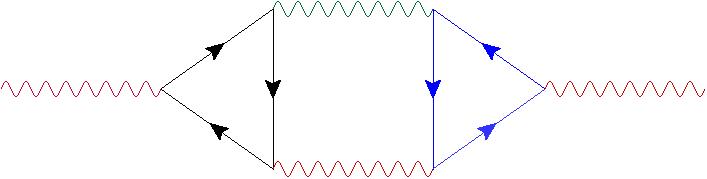} 
}
\subfigure[]{
\includegraphics[width=0.22 \textwidth]{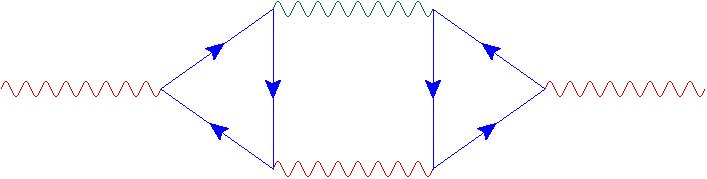} 
}
\subfigure[]{
\includegraphics[width=0.22 \textwidth]{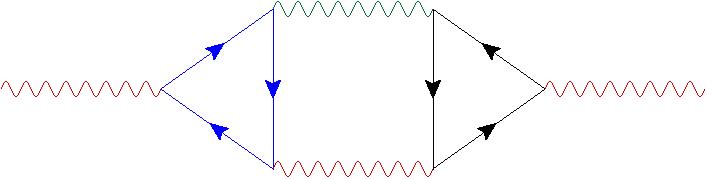} 
}
\subfigure[]{
\includegraphics[width=0.22 \textwidth]{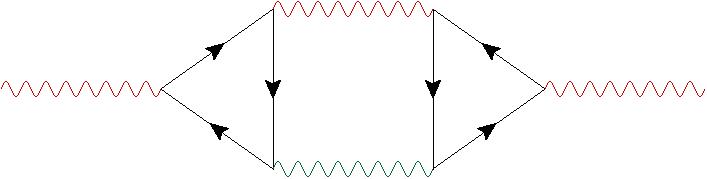} 
}
\subfigure[]{
\includegraphics[width=0.22 \textwidth]{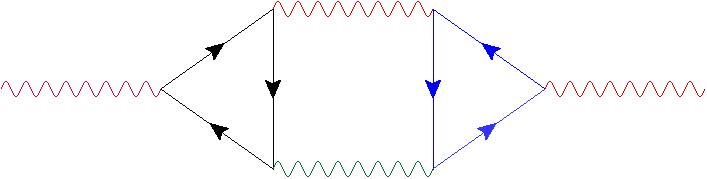} 
}
\subfigure[]{
\includegraphics[width=0.22 \textwidth]{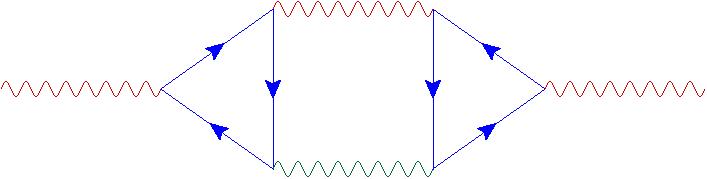} 
}
\subfigure[]{
\includegraphics[width=0.22 \textwidth]{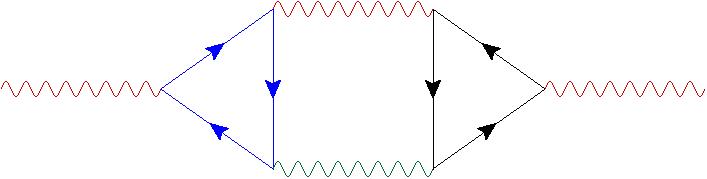} 
}
\end{center}
\caption{\label{ALdiag}
Contributions from the Aslamazov-Larkin diagrams in the particle-particle channels to the three-loop boson self-energy of the $a_c$ transverse gauge field. The red and green wavy lines denote the $a_c$ and $a_s$ propagators, respectively. The black and blue solid lines with arrows represent the $\psi_{1,\pm,j}$ and $\psi_{2,\pm,j}$ fermion propagators, respectively. All these sixteen diagrams give a total contribution proportional to $4\,{e}^2_c \left( \tilde{e}_c^2 
+\tilde{e}_s^2 \right)$.
}
\end{figure*}

Using the same arguments as the single gauge field case, we will have the following
nonzero $Z_\zeta^{(1)}$'s:
\begin{widetext}
\begin{align}
Z_{0}^{(1)} & =  
- \frac{ u_0 \left( \tilde{e}_c +{\tilde e}_s \right)} {N } 
  - \frac{ v_0 \left(  \tilde{e}_s +{\tilde e}_c\right)^2  } {N^2}\,,
 \quad 
 Z_{1}^{(1)}  = 
 -\frac{ u_1 \left( \tilde{e}_c +{\tilde e}_s \right) }   {N }
- \frac{ v_1 \left( \tilde{e}_c +{\tilde e}_s\right)^2 } {N^2 }\,,
\quad
Z_{2}^{(1)}  = 
- \frac{ w \left( \tilde{e}_c +{\tilde e}_s \right)^2 } {N^2}\,,\nn
Z_{4_s}^{(1)}  & = 
-\frac{  u_4  \left( \tilde{e}_c +{\tilde e}_s \right) } {N}
- \frac{ y \left( \tilde{e}_c +{\tilde e}_s \right)^2  } {N^2 }\,,
\quad
Z_{4_c}^{(1)} = 
-\frac{  u_4  \left( \tilde{e}_c +{\tilde e}_s \right) } {N}
- \frac{ y \left( \tilde{e}_c +{\tilde e}_s\right)^2  } {N^2}\,,
\end{align}
including the one- and two-loop corrections for $ m = 1 $.
This leads to the beta functions:
\begin{align}
& - \frac{\beta_{\tilde{e}_c}} {{\tilde e}_c} =
- \frac{\beta_{\tilde{e}_s}} {{\tilde e}_s}
\nn &
=    \frac{2  \left[ 2 \,u_1 \left( \tilde{e}_c +{\tilde e}_s\right) + 3 \,N\right]
  \epsilon}
 {9 \,N}
+
\frac{2 \left(2 \,u_0+u_1-4 \, u_4\right) \left( \tilde{e}_c +{\tilde e}_s\right) }
{9 N}
- \frac{4 \left[ 
-u_1^2-2 \,u_0 \,u_1 + 4 \,u_4 \,u_1
-3 \left(2 \, v_0 +  v_1 -3\, w \right)
+ 12 \,y \right ]
 \left( \tilde{e}_c +{\tilde e}_s\right)^2 }
{27\, N^2} \, ,
\end{align}
which again have a continuous line of fixed points defined by:
\begin{align}
\frac{{\tilde{e}_s}^*  + {\tilde e}_c^*}{N}
= \frac{3 \,\epsilon }{2\, u_0 + u_1-4 \,u_4}
-\frac{18
 \left(2 \,v_0+ v_1-3 \,w-4 \,y\right)
 \epsilon ^2 }
{\left(2 \,u_0+ u_1-4 \,u_4\right)^3} 
\nn & \qquad+
\mathcal{O} \big(\epsilon^3 \big) .
\end{align}
\end{widetext}

However, at three-loop order, there are diagrams giving rise to terms in the beta functions which are not proportional to $\left( \tilde{e}_s  + {\tilde e}_c\right)$ raised to some integer power. Some of them are shown in Fig.~\ref{ALdiag}, which are the Aslamazov-Larkin diagrams in the particle-particle channels contributing to the self-energy of the $a_c$ transverse gauge field. All these sixteen diagrams give a total contribution proportional to $4\,{e}_c^2 \left( \tilde{e}_c^2 
+\tilde{e}_s^2 \right)$. Similarly, the Aslamazov-Larkin diagrams in the particle-hole channels
will contribute with a term proportional to $4\,{e}^2_c \left( \tilde{e}_c^2 
+\tilde{e}_s^2 \right)$. One can also easily conclude that such corrections to the self-energy of the $a_s$ transverse gauge field will be proportional to $4\,{e}_s^2 \left( \tilde{e}_c^2 
+\tilde{e}_s^2 \right)$.
Note that if the two fermions had carried the same charge under the two gauge gauge fields, then the contributions would have been proportional to $4\,{e}^2_c \left( \tilde{e}_c 
+\tilde{e}_s \right)^2$ and $4\,{e}^2_s \left( \tilde{e}_c 
+\tilde{e}_s \right)^2$, respectively, for the $a_c$ and $a_s$ fields, leading to the preservation of the fixed line feature.

For $m>1\,,$ the UV/IR mixing will render the higher-loop corrections to be $k_F$-suppressed
and hence they will have no effect on the fixed line.
Therefore, the fixed line feature is generically not altered by going to higher loops.

\section{\label{conclude}Conclusion}

In this paper, we have applied the dimensional regularization scheme, developed for non-Fermi liquids arising
at Ising-nematic quantum critical point, to the case of non-Fermi liquids arising from transverse gauge field couplings with finite-density fermions. This has allowed us to access the interacting fixed points perturbatively in an expansion in $\epsilon\,,$ which is the difference between the upper critical dimension ($d_c =m+\frac{3}{m+1}$) and the actual physical dimension ($d_{\text{phys}}=m+1$) of the theory, for a Fermi surface of dimension $m$. We have extracted the scaling behaviour for the case of one and two $U(1)$ gauge fields.

There is a crucial difference in the matrix structure of the couplings in the cases of Ising-nematic order parameter and gauge fields. This arises from the fact that the fermions on the antipodal points of the Fermi surface couple to the Ising-nematic order parameter (transverse gauge) with the same (opposite) sign(s). 
Hence, although we get the same values of critical dimension and critical exponents, the differences will show up in the renormalization of some physical quantities like the $2 k_F$ scattering (backscattering involving an operator that carries a momentum $2k_F$) amplitudes. This operator can also be identified as the CDW instability. In particular, we have found that this CDW ordering is enhanced in the vicinity of the non-Fermi liquid critical point for tranverse gauge field(s) for $m=1$, in contrast to the Ising-nematic scenario.

The $U(1) \times U(1)$ is particularly interesting in the context of recent works which show that this scenario is useful to describe the phenomena of deconfined Mott transition, and
deconfined metal-metal transition \cite{debanjan}. In Ref.~\cite{debanjan}, Zou and Chowdhury found that in $(2+1)$ spacetime dimensions and at one-loop order,
these systems exhibited a continuous line of stable fixed points, rather than a single one. Their method involved modifying the bosonic dispersion (such that it becomes nonanalytic in the momentum space), and then carrying out a double expansion in two small parameters \cite{nayak1,mross}. Our method avoids this issue by employing the dimensional regularization scheme. We also have the advantage that we could analyze a critical Fermi surface of generic dimensions, and also perform higher-loop diagrams giving order by order corrections in $\epsilon$. The discovery of a fixed line for the $U(1) \times U(1)$ theory in Ref.~\cite{debanjan} raised the question whether this feature survives when we consider either higher dimensions or higher loops. Our computations show that definitely higher dimensions do not reduce the fixed line to discrete fixed points (or no fixed point at all).
Regarding higher-loop corrections, we have not performed those explicitly, but through arguments based on the previous results for the Ising-nematic critical points
\cite{Lee-Dalid,ips-uv-ir1,ips-uv-ir2}, we have predicted that although the two-loop corrections will not affect the fixed line, the presence of certain three-loop diagrams will. If the fixed line feature is destroyed by three-loop and / or higher order corrections for $m=1$, this will have the following possibilities for Ref.~\cite{debanjan}: (1) The fixed line degenerates into fixed points, which might be stable or unstable.
(2) The zeros of the beta functions have no finite solution (i.e. no fixed point exists).
The physical consequences of these scenarios have been explained in detail in Sec.~V of Ref.~\cite{debanjan}.

In the future, it will be worthwhile to carry out this entire procedure for the case of $SU(2)$
gauge fields \cite{debanjan}. It is also possible that the non-Fermi liquid fixed point/line for $m=1$ is masked by a CDW phase. Extending our RG analysis to a situation where fluctuations due to the transverse gauge field(s) and CDW are treated on equal footing would be an interesting problem for future study. Another direction is to compute the RG flows for superconducting instabilities in the presence of the transverse gauge field(s), as was done in Ref.~\cite{ips-sc} for the Ising-nematic order parameter.


\begin{acknowledgments}
We thank Debanjan Chowdhury for stimulating discussions. We are especially grateful to Andres Schlief
for valuable comments on the manuscript. We also thank Sung-Sik Lee for pointing out that Aslamasov-Larkin diagrams will affect the fixed line feature at three-loop order.
\end{acknowledgments}


\begin{widetext}
\appendix

\section{Computation of the Feynman diagrams at one-loop order}
\label{app:oneloop}

\subsection{One-loop boson self-energy}
\label{app:oneloopbos}

In this subsection, we compute the one-loop boson self-energy:
\begin{align}
\label{bosloop}
\Pi_1 (q) = & -e^2 \mu^x
\int dk\,\text{Tr}
\left[ \gamma_{0}\, G_0 (k+q)\,\gamma_{0}\, G_0 (k) \right ]
=  
2 \, e^2  \mu^x \int dk\, 
\frac{ k_0 \left(k_0+q_0 \right)
 -\tilde{\vec K} \cdot \left ( \tilde{\vec K} + \tilde{\vec Q} \right ) - \delta_q \,\delta_{k+q}}
{[\vec K^2 + \delta_k^2]\, [(\vec K +\vec Q)^2 + \delta_{k+q}^2 ]} \, 
 e^{-\frac{{\vec{L}}_{(k)}^2 + {\vec{L}}_{(k+q)}^2} { \mu \, {\tilde{k}}_F }}\,.
\end{align}

We first integrate over $k_{d-m}$ to obtain
\footnote{While performing the integral for $k_{d-m+1}$, we can neglect the contribution from the exponential term as $k_{d-m+1}$ already appears in the denominator and is appropriately damped. For extracting leading order singular behaviour in $k_F$, this is sufficient.}: 
\begin{align}
\Pi_1 (q) &=  e^2 \, \mu^x \int \frac{ d{\vec{L}}_{(k)} \, d\vec K}{(2\,\pi)^d} 
\frac{ 2\,k_0 \left(k_0+q_0 \right) -\vec K \cdot (\vec K +\vec Q)}
{ |\vec K|\,|\vec K +\vec Q|\, 
\left[ \big (\, \delta_q +2  \, {\vec{L}}_{(q)}^i \, {\vec{L}}_{(k)}^i \, \big )^2 + \big(\, |\vec K +\vec Q|+|\vec K|\, \big)^2
\right] } \big(\,|\vec K +\vec Q|+|\vec K| \, \big) 
\, e^{-\frac{{\vec{L}}_{(k)}^2 + {\vec{L}}_{(k+q)}^2} { \mu \, {\tilde{k}}_F }}    \nn
& \quad -
e^2 \, \mu^x \int \frac{ d{\vec{L}}_{(k)} \, d\vec K}{(2\,\pi)^d} 
\frac{ |\vec K +\vec Q|+|\vec K| }
{  \big (\, \delta_q +2  \, {\vec{L}}_{(q)}^i \, {\vec{L}}_{(k)}^i \, \big )^2 + \big(\, |\vec K +\vec Q|+|\vec K|\, \big)^2 }  
\times  e^{-\frac{{\vec{L}}_{(k)}^2 + {\vec{L}}_{(k+q)}^2} { \mu \, {\tilde{k}}_F }}  \,,
\end{align}
where we have chosen the coordinate system such that ${\vec{L}}_{(q)} = (q_{d-m+1},0, 0,\ldots,0) $. Since the problem is rotationally invariant in these directions and $\Pi_1(q)$ depends only on the magnitude of ${\vec{L}}_{(q)}$, the final result is independent of this choice.

Making a change of variable, 
$u=\delta_q +2 \, q_{d-m+1} \, k_{d-m+1} \,,$
and integrating over $u$,  we get:
\begin{align}
I &\equiv  \int \frac{d k_{d-m+1}}{2 \,\pi} \frac{1} { \big (\, \delta_q +2 \, q_{d-m+1} \, k_{d-m+1} \, \big )^2 + \big(\, |\vec K +\vec Q|+|\vec K|\, \big)^2 }
= \frac{1}{4 \,  |\vec{L}_{(q)}| \,  \big(\, |\vec K +\vec Q|+|\vec K|\, \big)} \,.
\end{align}
The rest of $\vec{L}_{(k)}$-integrals evaluate to $J^{m-1}\,,$ where
\begin{align}
\label{eqJ}
J \equiv \int_{-\infty}^{\infty} \frac{dy}{2 \pi} \, \exp{ \Big \lbrace \frac{-2y^2} { \mu \, {\tilde{k}}_F } \Big \rbrace} =\sqrt{\frac{ \mu \, {\tilde{k}}_F }{8 \pi}} \,.
\end{align}
Hence the self-energy expression reduces to:
\begin{align}
\label{pia11}
\Pi_1 (q) =
\frac{   e^2 \, \mu^x } {  2^{m+1} \, |\vec{L}_{(q)}|}  \Big ( \frac{ \mu \, {\tilde{k}}_F } {2 \pi} 
\Big )^{\frac{m-1}{2}} \,  I_1 (d-m, \vec Q)\,, 
\end{align}
where
\begin{align}
I_1 (d-m, \vec Q) 
&= \int \frac{  d\vec K}{(2\,\pi)^{d-m} } 
\left [\frac{k_0 \left(k_0+q_0 \right) -\tilde{\vec K } \cdot \left ( \tilde{\vec K} + \tilde{\vec Q} \right )}  {|\vec K +\vec Q| \,|\vec K|} 
- 1  \right ]   .
\end{align}
The ($d-m$)-dimensional integral in $I_1 (d-m,\vec Q)$ can be done 
using the Feynman parametrization formula
\begin{align}
\label{feynm}
\frac{1}{A^{\alpha} \,B^{\beta}}= 
\frac{\Gamma (\alpha +\beta)}{\Gamma (\alpha) \,\Gamma (\beta)}
\int_0^1 \,dt\, \frac{t^{\alpha-1}\,(1-t)^{\beta-1}}
{\left[ t \,A +(1-t) \,B\right]^{\alpha+\beta}} \,.
\end{align}
Substituting $\alpha=\beta=1/2$, 
$A=| \vec K + \vec Q|^2 $ and $B=|\vec K|^2 $, we get:
\begin{align}
I_1 (d-m, \vec Q) = \frac{1}
{ \pi \, (2 \pi)^{d-m}  } 
\int_0^1 \frac{dt }{\sqrt{ t \, (1-t)}}\,
\int {d \vec K} \left[
\frac{ k_0 \left(k_0+q_0 \right) 
- \tilde{\vec K} \cdot \left (\tilde{\vec K}+\tilde{\vec Q} \right )}
{   x \, |\vec K  + \vec Q|^2 +  (1-t) \vec K^2 } 
-1 \right ] .
\end{align}
Introducing the new variable $ \vec u = \vec K + t \, \vec Q \,,$ $I_1$ reduces to:
\begin{align}
 I_1(d-m, \vec Q) 
= \frac{1}
{\pi \, (2 \,\pi)^{d-m}  } 
\int_0^1 \frac{dt }  {\sqrt{ t \, (1-t)}}\,
\int {d^{d-m}\vec u} \,
\left [
\frac{ 2 \left \lbrace  u_0^2 - t \,(1-t) \, q_0^2  \right \rbrace  
-2\,\vec u^2  }
{   \vec u ^2 + t \,(1-t) \,\vec Q^2}
 \right ].
\end{align}
Again, we use another new variable $\vec v$, defined by $\vec u = \sqrt{t\,(1-t)} \,\vec v \,,$ so that
\begin{align}
\label{i1}
I_1 (d-m, \vec Q)
&= -\frac{2^{-2 d+2 m+1} \,
\pi ^{-d+m+\frac{1}{2}} \Gamma \left(\frac{d-m+1}{2} \right)}{\Gamma \left(\frac{d-m+2}{2} \right)}
\int {d^{d-m}\vec v} \,
\frac{    q_0^2   + \tilde{ \vec v}^2  }
{   \vec v ^2 + \vec Q^2} \,. 
\end{align}
Using 
\begin{align}
\int_0^{\infty}\, dy \, \frac{ y^{n_1}} {( y^2 + C)^{n_2}} =
\frac{ \Gamma  \left (\frac{n_1+1}{2} \right ) \,  \Gamma \left (n_2-\frac{n_1+1}{2} \right ) }
{2 \, \Gamma(n_2)}  
\, C^{\frac{n_1+1}{2}-n_2} \,,
\end{align}
 and the volume of the $(n-1)$-sphere (at the boundary of the $n$-ball of unit radius)
\begin{align}
S^{n-1} \equiv \int  d \Omega_n = \frac{2 \,\pi^{n/2}}  { \Gamma \left (n/2  \right )} \,,
\end{align}
we finally obtain the one-loop boson self-energy to be:
\begin{align}
\label{api}
\Pi_1 (k) 
& =
-\frac{ \beta(d,m)\,  e^2 \, \mu^x } {   |\vec{L}_{(q)}|}  \Big (  \mu \, {\tilde{k}}_F  \Big )^{\frac{m-1}{2}} 
 \left[ k_0^2 + ( m+1-d)\,{\tilde {\vec K}}^2 \right] \,|\vec K|^{d-m-2}
 \,,
\end{align}
with
\begin{align}
\beta(d,m)
 &= \frac{ 1 } {  2^{m+1}}  \Big ( \frac{1 } {2 \pi} 
\Big )^{\frac{m-1}{2}}
\frac{2^{-2 d+2 m+1}\, \pi ^{\frac{-d+m+3}{2} } 
\,\Gamma (d-m) \,\Gamma (m+1-d) }
{\Gamma ^2 \left(\frac{ d-m+2} {2}\right) \Gamma \left(\frac{m+1-d}{2} \right)}
=
\frac{2^{\frac{1+m-4d}{2} }\, \pi ^{\frac{4-d}{2}} 
\,\Gamma (d-m) \,\Gamma (m+1-d) }
 {\Gamma ^2 \left(\frac{ d-m+2} {2}\right) \Gamma \left(\frac{m+1-d}{2} \right)} \,.
\end{align}

\subsection{One-loop fermion self-energy}
\label{app:oneloopfer}

Here we compute the one-loop fermion self-energy $\Sigma_1 (q) $
by using the dressed propagator for boson
which includes the one-loop self-energy $\Pi_1(k)$:
\begin{align}
\label{sigma1int}
\Sigma_1 (q) &= \frac{ e^2  \, \mu^{x}}  {N} 
\int dk \,
\gamma_{0} \, G_0 (k+q)\, \gamma_{0} \,D_1 (k) 
=  \frac{ \mathrm{i}\,e^2 \,  \mu^{x}}{N}  \int dk\,
D_1 (k) \frac{ \tilde{\vec \Gamma } \cdot \left( \tilde{\vec K } 
+ \tilde{\vec Q }  \right ) + \gamma_{d-m} \, \delta_{k+q} - \gamma_0\left(  k_0+q_0 \right)}
{\left ({\vec K } + {\vec Q} \right )^2 +\delta_{k+q}^2}   \, 
 e^{-\frac{{\vec{L}}_{(k+q)}^2} { \mu \, {\tilde{k}}_F }} \, .
\end{align}
 Integrating over 
$k_{d-m}\,,$ we get:
\begin{align}
\Sigma_1 (q) = \frac{\mathrm{i}\,  \,e^2 \,  \mu^{x}}{ 2 N}  \int \frac{d ^{d} k}{(2\pi)^{d}} \, 
D_1 (k) \frac{ \tilde{\vec \Gamma } \cdot \left( \tilde{\vec K } + \tilde{\vec Q }  \right ) 
-\gamma_0\left(  k_0+q_0 \right) 
}
{|\vec K +\vec Q|}   \, \times e^{-\frac{{\vec{L}}_{(k+q)}^2} { \mu \, {\tilde{k}}_F }} \, .
\end{align}
Since only $D_1 (k) \, e^{-\frac{{\vec{L}}_{(k+q)}^2} {k_F}} $ depends on $ {\vec{L}}_{(k)}$, let us first perform the integral:
\begin{align}
\label{i2}
I_2(k) & \equiv \int   \frac{d  {\vec{L}}_{(k)} }  {(2\pi)^{m}} \, \frac{ e^{-\frac{{\vec{L}}_{(q+k)}^2} {k_F}}}
 { {\vec{L}}_{(k)}^2  +  \beta(d,m) \, e^2 \, \mu^{x} \,( \mu \, {\tilde{k}}_F )^{ \frac{m-1}{2}}  \, \frac{  |\vec K|^{d-m}}{ |\vec{L}_{(k)}| }
 \times   \left[ d-m-1  +(m-d)\frac{ k_0^2}{|\vec K|^2}\right]} 
\nn & 
= \frac{  \pi^{\frac{2-m}{2}} }
{3 \times 2^{m-1} \, \Gamma{(m/2)} \,  | \sin \lbrace (m+1) \pi/3 \rbrace | 
\left [ \beta(d,m) \, e^2 \, \mu^{x} \,( \mu \, {\tilde{k}}_F )^{ \frac{m-1}{2}}  \,  |\vec K|^{d-m-2} 
\times   \left \lbrace   k_0^2 + ( m+1-d)\,{\tilde {\vec K}}^2 \right  \rbrace 
 \right ]^{\frac{2-m}{3}} } \,.
\end{align} 

Now the expression for the self-energy can be written as:
\begin{align}
\Sigma_1 (q) = \frac{ \mathrm{i} \, e^{2(m+1)/3} \, \mu^{ x \, (m+1)/3} \, \pi^{\frac{2-m}{2}}
 \times I_3 (d-m, \vec Q)}
{6 N 
\times 2^{m-1} \, \Gamma{(m/2)} \,  |\sin \left(  \frac{m+1}{3} \,\pi \right )|\,
\left[ \beta(d,m) \right] ^{\frac{2-m} {3}}  \, ( \mu \, {\tilde{k}}_F ) ^{(m-1)(2-m)/6} }  \,,
\end{align}
where
\begin{align}
& I_3 (d-m, \vec Q) 
\nn & = \int \frac{ d \vec K}{(2\pi)^{d-m}}
\frac{ \tilde{\vec \Gamma } \cdot \left( \tilde{\vec K } + \tilde{\vec Q }  \right )
- \gamma_0\left(  k_0+q_0 \right) 
}
{ \left [  \left \lbrace  k_0^2 + ( m+1-d)\,{\tilde {\vec K}}^2 \right \rbrace \,|\vec K|^{d-m-2}
 \right ]^{\frac{2-m}{3}}
\,|\vec K +\vec Q|}
\nn  & =  \frac{\Gamma \left(  \frac{ 2 d+m^2+3 - m\,(d+2) }  {6}  \right)  }
{\Gamma \left(\frac{1}{2}\right) \Gamma \left(\frac{2-m}{3}\right) 
\Gamma \left(\frac{(d-m-2) (2-m)}{6} \right)}
 \int_0^1  dt_1 \int_0^{1-t_1} dt_2\,
 \frac{
(1-t_1-t_2)^{ \frac{(d-m-2)\,(2-m)}{6}-1}  }
{t_1^{\frac{m+1}{3}} \,\sqrt{t_2} }
\nn &   \quad \times
 \int \frac{ d \vec K}{(2\pi)^{d-m}}
\frac{ \tilde{\vec \Gamma } \cdot \left(\tilde{\vec K } + \tilde{\vec Q }\right )
-\gamma_0  \left(  1 -t_2\right) q_0 
}
{  \left [  
t_1\,  (m+1-d)\,{\tilde{\vec K}}^2 
 +  t_2 \left(\tilde{\vec K }+ \tilde{\vec Q }\right)^2
+ (1-t_1-t_2)\,  \tilde{\vec K}^2 + q_0^2\, t_2\, (1 - t_2)+ k_0^2
 \right ]^{\frac{ 2 d+m^2+3 - m\,(d+2) }  {6}}} 
 \nn & =
 \frac{\pi ^{\frac{m-d-1}{2} } \, \Gamma \left(\frac{ 3-(d-m)\,(m+1)} {6} \right)}
 {2^{d-m}   \,\Gamma \left(\frac{2-m}{3}\right) \Gamma \left(\frac{ m^2-d (m-2)-4}{6} \right)}
\int_0^1  dt_1 \int_0^{1-t_1} dt_2\Bigg [
 \frac{
(1-t_1-t_2)^{ \frac{(d-m-2)\,(2-m)}{6}-1}  }
{t_1^{\frac{m+1}{3}} \,\sqrt{t_2} } 
\,  \frac{ \gamma_0  \left(  1 -t_2\right) q_0 
- \left( \tilde{\vec \Gamma } \cdot \tilde{\vec Q } \right )
\left [\frac{ \left(  1+m\,t_1- d\,t_1 -t_2\right)}
{1+m\,t_1- d\,t_1}\right ] }
{  \left ( 1+m\,t_1- d\,t_1   \right )^{ \frac{(d-m)\, (2-m)}{6}  }
}
\nn & \quad
\left \lbrace \frac{\tilde{\vec Q}^2\,t_2  \left( 1+m\,t_1- d\,t_1 -t_2 \right) } 
{ \left( 1+m\,t_1- d\,t_1  \right)^2 } +\frac{ q_0^2\, t_2\, (1 - t_2)} {1+m\,t_1- d\,t_1}  
 \right \rbrace^{\frac { (d-m)\,(m+1)-3} {6}  } 
\left \lbrace \Theta \left(  1+m\,t_1- d\,t_1  \right)
 + fac \times \Theta \left(  d\,t_1-1-m\,t_1  \right) \right \rbrace
 \Bigg]
 \,.
\end{align}
where $fac =(-1)^{\frac{(m-2) \,(d-m)}{3} } 
\mathrm{i}^{d-m+1} 
\left[ (-1)^{\frac{1}{6} (m-2) (m-d)} \cos \left(\frac{ (m+1) (d-m)}{6} \pi \right)
-\cos \left(\frac{d-m}{2} \pi  \right)\right ] \csc \left(\frac{(m-2) (d-m)}{6} \pi  \right) \,.$
From this expression, it is clear that $I_3 (d-m, \vec Q)$ blows up when the argument of the gamma function in the numerator blows up, \textit{i.e.} $ \frac{3-(d-m)\,(m+1)} {6}  =0\,.$
This implies that $\Sigma_1 (q) $ blows up logarithmically in $\Lambda$ at the critical dimension
\begin{align}
d_c(m) =m+\frac{3}{m+1}\,.
\end{align}
The integrals over $t_1$ and $t_2$ are convergent, but their values have to be computed numerically for a given $m$.

Expanding in $\epsilon $ defined as $ d=m+\frac{3}{m+1}-\epsilon$, we obtain:
\begin{align}
\Sigma_1(q) =
-\frac{ \mathrm{i} \, 
e^{\frac{2\,(m+1)} {3} } 
\left[ u_0  \left(  \frac{\mu}{ |q_0| } \right)^{ \frac{m+1}{3} \epsilon}
\gamma_0\,q_0 
+u_1  \left(  \frac{\mu}{ |\tilde{\vec Q}| } \right)^{ \frac{m+1}{3} \epsilon}
\left( {\tilde{\vec \Gamma}} \cdot {\tilde{\vec Q}} \right)
 \right]  }
{N \, {\tilde{k}}_F ^{ \frac{(m-1)(2-m) } {6}}
\,\epsilon}
+\text{ finite terms}  \,,
\end{align}
where $u_0, \, u_1 \geq 0\,.$
Numerically, we find:
\begin{align}
\begin{cases}
u_0 = 0.0201044 \,, \quad u_1 =1.85988 & \text{ for } m=1 \\
 u_0 = u_1 = 0.0229392 & \text{ for } m=2
\end{cases} \,.
\end{align}

\subsection{One-loop vertex correction}
\label{oneloopvert}

In general, the one-loop fermion-boson vertex function 
$\Gamma_1 (k,q)$ depends on both $k$ and $q$. 
In order to extract the leading $1/\epsilon$ 
divergence, however, it is enough to look at the $q \rightarrow 0$ limit. In this limit, we get:
\begin{align}
\Gamma_1 (k,0)&= \frac{ e^2 \,\mu^{x}}{N} 
\int \frac{d^{d+1} q}{(2\pi)^{d+1}} \, \gamma_{0} \, G_0 (q)\,
\gamma_{0} \, G_0 (q) \,  \gamma_{0} \, D_1 (q-k)\nn
&= \frac{e^2 \,\mu^{x}}{N}
\int \frac{d^{d+1} q} {(2\pi)^{d+1}} D_1 (q-k) \, \gamma_{0}\, 
\left[ 
\frac{ 1 } { \vec Q^2  + q_{d-m}^2  }
- \frac{ 2\,q_0^2 } 
{\left( q_0^2+\tilde{\vec Q}^2  + q_{d-m}^2 \right)^2 }
\right] e^{-\frac{2 \, \vec{L}_{(q)}^2}{\mu \tilde{k}_F}} 
\nn &=
 \frac{e^2 \,\mu^{x}}{2\,N}
\int \frac{ d{\vec Q}
\,d {\vec{L}}_{(q)} } {(2\pi)^{d}} D_1 (q) \, \gamma_{0}\,
  \frac{ (\tilde{\vec Q}+ \tilde{\vec K} )^2}
  {\left [ (q_0+k_0^2 )^2+ (\tilde{\vec Q} +\tilde{\vec K})^2\right ]^{3/2}}
e^{-\frac{2 \, \vec{L}_{(q+k)}^2}{\mu \tilde{k}_F}} \,.
\end{align}

Using Eq.~(\ref{i2}), the above expression reduces to:
\begin{align}
\Gamma_1 (k,0)&= 
\frac{e^2 \,\mu^{x}\, \gamma_{0}} {N}
\frac{  \pi^{\frac{m}{2}+1-d} }
{3 \times 2^{d} \, \Gamma{(m/2)} \,  | \sin \left(  \frac{m+1}{3} \,\pi \right ) | 
\left [ \beta(d,m) \, e^2 \, \mu^{x} \,( \mu \, {\tilde{k}}_F )^{ \frac{m-1}{2}}  \right ]^{\frac{2-m}{3}} } 
\times I_4(d,m)\,,
\end{align}
where
\begin{align}
&  I_4(d,m)
\nn &  = \int \frac{d{\vec Q}}
{\left [   |\vec Q|^{d-m-2}   \left \lbrace   q_0^2 + ( m+1-d)\,{\tilde {\vec Q}}^2 \right  \rbrace 
 \right ]^{\frac{2-m}{3}}  }
\frac{ (\tilde{\vec Q}+ \tilde{\vec K} )^2}
  {\left [ (q_0+k_0^2 )^2+ (\tilde{\vec Q} +\tilde{\vec K})^2\right ]^{3/2}}
\nn & = \int_0^1  dt_1 \int_0^{1-t_1} dt_2\,
 \frac{\Gamma \left( \frac{ 2 d+m^2+9-m\,(d+2)}{6}  \right)
t_1^{-\frac{m+1}{3}} \,t_2^{\frac{1}{2}} \,(1-t_1-t_2)^{ \frac{(d-m-2)\,(2-m)}{6}-1}  }
{\Gamma \left(\frac{3}{2}\right) \Gamma \left(\frac{2-m}{3}\right) 
\Gamma \left(\frac{(d-m-2) (2-m)}{6} \right)}
\nn &   \quad \times
 \int \frac{ d \vec Q}{(2\pi)^{d-m}}
\frac{ (\tilde{\vec Q}+ \tilde{\vec K} )^2}
{  \left [  
t_1\,  (m+1-d)\,{\tilde{\vec Q}}^2 
 +  t_2\left(\tilde{\vec Q }+ \tilde{\vec K }\right)^2
+ (1-t_1-t_2)\,  \tilde{\vec Q }^2 + k_0^2\, t_2\, (1 - t_2)+ q_0^2
 \right ]^{\frac{-(d+2) m+2 d+m^2+9}{6} }}  
\nn & = \int_0^1  dt_1 \int_0^{1-t_1} dt_2\,
 \frac{
t_1^{-\frac{m+1}{3}} \,t_2^{\frac{1}{2}} \,(1-t_1-t_2)^{ \frac{(d-m-2)\,(2-m)}{6}-1}  }
{\Gamma \left(\frac{3}{2}\right) \Gamma \left(\frac{2-m}{3}\right) 
\Gamma \left(\frac{(d-m-2) (2-m)}{6} \right)} 
\times
\frac{ \Gamma \left(\frac{ (d-m)\, (2-m)} {6} + 1\right)}
{ 
\left (1+ m\,t_1- d\,t_1 \right )^{ \frac{ (d-m) (m-2)-6 }  {6}  } 
}
\times  J_1 \,.
\end{align}
Here,
\begin{align}
 & J_1 = \int \frac{ d \tilde{\vec Q}}{(2\pi)^{d-m}}
\frac{ \tilde{\vec Q}^2+  A^2  }
{  \left [{\tilde{\vec Q}} ^2+ B\right ] ^{ \frac{ (d-m) (2-m)+6 }  {6}  } } \nn
& =
\frac{ B^{\frac{ -9+d-m^2+m(d-1)}{6} } 
\, \Gamma \left(\frac{(m-d)\, (m+1)+3} {6} \right) 
}
{3 \times 2^{d-m+1}\, \pi ^{\frac{d-m+1}{2} } \,\Gamma \left(\frac{ (d-m) (2-m)}{6}+1\right)}
\left[ \left \lbrace (m-d)\, (m+1)+3\right \rbrace  A^2+3 \, (d-m-1) \,B
\right]\,,
\end{align}
where $B \equiv   \frac{\tilde{\vec K}^2\,t_2  \left( 1+m\,t_1- d\,t_1 -t_2 \right) } 
{ \left( 1+m\, t_1- d\,t_1  \right)^2 } +\frac{ k_0^2\, t_2\, (1 - t_2)} {1+m\,t_1- d\,t_1}  $
and $ A^2 =  \left( \frac{ 1+m\,t_1- d\,t_1 -t_2}
{1+m \,t_1 -d \,t_1} \right )^2 \tilde{\vec K} ^2   \,.$
Expanding in $\epsilon = m+\frac{ 3}{m+1}- d$, we obtain:
\begin{align}
 \Gamma_{1}(k,0)=
- \frac{e^{\frac{2 \,(m+1)}{3}}\,
 \,\gamma_{0}}
    {N \,\tilde{k}_{F}^{ (m-1) \, (2-m) / 6}\,\epsilon}\, u_4 \,
 \left( \frac{\mu} { | \tilde{\vec K}|} \right)^{\frac{(m+1) \,\epsilon}{3}}  
 \left[  {\mathcal F}\bigg ( \frac{|k_0|} { | \tilde{\vec K}|} \bigg) \right]^\epsilon
    +\text{finite terms} \,,
\end{align}
where $u_4 \geq 0$ and ${\mathcal F}$ is some dimensionless function of ${|k_0|} /{ | \tilde{\vec K}|}$. Specifically, we have:
\begin{align}
 u_4 = \begin{cases}
0.0000706373 & \text{ for } m=1 \\
    0 &\text{ for } m=2
\end{cases} \,.  
\end{align}

\subsection{Renormalization of the $2k_F$ scattering amplitude}
\label{oneloop2kf}

The one-loop corrections to
the $2k_F$ scattering amplitude $g_{2k_F}$, captured by Fig.~\ref{fig:2kf}, can be obtained as: 
\begin{align}
 \frac{ \mathrm{i}\,e^2 \,\mu^x\,g_{2k_F}}{N} 
\int dq  \,\gamma_{0}^T \,G_0^{T} (q) \,\gamma_0  \,
G_0 (-q) \,\gamma_{0} \,D_1 (q-k)\,,
\end{align}
where the superscript $T$ denotes transpose of the corresponding matrix.
If $d-m=2 $, we have 
$\gamma_0^{T} = -\sigma_y = -\gamma_0$, 
$\gamma_1^{T}= \sigma_z = \gamma_1$ and 
$\gamma_2^T =\sigma_x = \gamma_2$.
For $2<d<5/2$, we generalize this as:
\begin{align}
\gamma_0^{T} &= -\gamma_0 \, , \quad
\gamma_\chi^T = \gamma_\chi\, ,
\text{ for  $\chi =1,\ldots,d-m$}. 
\end{align}
Using this, we obtain the one-loop correction
\begin{align}
& \delta g_{2k_F}^{(1)}=
\frac{ \mathrm{i}\, e^2 \,\mu^x\,g_{2k_F}}{N} 
\int dq  \,\gamma_{0} \,\frac{{\mathbf \Gamma}^T \cdot \mathbf Q +\gamma_{0}^T\,\delta_{q}}
{ \mathbf Q^2  +\delta_{q}^2} \,\gamma_0  \,
\frac{-{\mathbf \Gamma} \cdot \mathbf Q +\gamma_{d-m}\,\delta_{-q}}
{ \mathbf Q^2  +\delta_{-q}^2}
\,\gamma_{d-m} \,D_1 (q-k) \nn
& = - \frac{ \mathrm{i}\,e^2 \,\mu^x\,g_{2k_F}}{N} 
\int dq  \,\gamma_{0} \,\frac{
q_0^2 + {\tilde {\vec Q}}^2 + \delta_q\,\delta_{-q}
}
{ \left(  \mathbf Q^2  +\delta_{q}^2 \right) 
\left( \mathbf Q^2  +\delta_{-q}^2  \right) } \,D_1 (q-k) 
+ \text{ terms irrelevant for } 2k_F \text{ scattering}\nn
&= -\frac{ \mathrm{i}\,e^2 \,\mu^x\,g_{2k_F}}{2\,N} 
\int \frac{d\mathbf Q\, d{{\mathbf L}_{(q)}}}  {(2\,\pi)^d}  \,
\frac{  \gamma_0\, |\mathbf Q +\mathbf K|   }
{    \left ( \mathbf Q + \mathbf K \right )^2  + {\mathbf L}_{(q+k)}^4  }  
\, \frac{ e^{-\frac{{\vec{L}}_{(q)}^2} {k_F}}}
 { {\vec{L}}_{(q)}^2  -  \beta(d,m) \, 
 e^2 \, \mu^{x} \,( \mu \, {\tilde{k}}_F )^{ \frac{m-1}{2}}  
 \, \frac{  |\vec Q|^{d-m-2}}
 { |\vec{L}_{(q)}| }
  \left[ (m+1-d)\,|\tilde{\vec Q}|^2 + q_0^2 \right]}\,.
\end{align}
In order to extract the divergent part, we can set $ {\mathbf L}_{(k)}= 0$, and scale ${\mathbf L}_{(q)}$ as $ \left[ e^2 \, \mu^{x} \,( \mu \, {\tilde{k}}_F )^{ \frac{m-1}{2}} \right ]^{1/3} 
{\mathbf L}_{(q)}$ (in addition to setting ${\mathbf L}_{(q)}$ to zero everywhere except $D_1(q)$) for the leading order behaviour in $\tilde e$. This finally gives the expression:
\begin{align}
\delta g_{2k_F}^{(1)} 
& =- \frac{ \mathrm{i}\,e^2 \,\mu^x\,g_{2k_F}}{2\,N} 
\int \frac{d\mathbf Q\, d{{\mathbf L}_{(q)}}}  {(2\,\pi)^d}  \,
\frac{  \gamma_0  }
{ |\mathbf Q +  \mathbf K |   }  
\, \frac{ |\vec{L}_{(q)}|}
 { |\vec{L}_{(q)}|^3  -  \beta(d,m) \,  |\vec Q|^{d-m-2}
   \left[ (m+1-d)\,|\tilde{\vec Q}|^2 + \left(q_0+k_0\right)^2 \right]}
\frac{1}
{\left[ e^2 \, \mu^{x} \,( \mu \, {\tilde{k}}_F )^{ \frac{m-1}{2}} \right ]^{\frac{2-m}{3}} } \nn
& =  \begin{cases}
\frac{ \mathrm{i}\, e^{\frac{4}{3}} 
\left( \frac{\mu} {|k_0|}  \right)^{\frac{2\,\epsilon}{3}}
\,g_{2k_F}}{ N} 
\times \frac{ - 0.0774559 \,\gamma_0 } {\epsilon}
& \text{ for } m=1 \\
0
& \text{ for } m=2
\end{cases} \,,
\end{align}
modulo terms irrelevant for the renormalization of $ 2k_F $ scattering.

\end{widetext}

\bibliography{biblio}

\end{document}